\documentclass[prb,aps,amsmath,amssymb,amsfonts,floatfix,superscriptaddress,showpacs,twocolumn]{revtex4}
\usepackage{longtable}
\usepackage{graphicx}
\usepackage{epsfig}
\usepackage[dvips]{color}
\usepackage{dcolumn}
\usepackage{nicefrac}

\newcommand{\bR}{\mathbf{R}}
\newcommand{\bK}{\mathbf{K}}
\newcommand{\bH}{\mathbf{H}}

\def\beq{\begin{equation}}
\def\eeq{\end{equation}}
\def\be{\begin{equation}}
\def\ee{\end{equation}}

\def\t{\mbox{tr}\,}
\def\cG0{{\cal G}_0}
\def\cG{{\cal G}}

\newcommand{\up}{\uparrow}
\newcommand{\don}{\downarrow}
\def\efat{\mbox{\boldmath$\varepsilon$}}
\def\a{\alpha}

\def\uc2{$U_{c2}$}
\def\uc1{$U_{c1}$}

\def\fd{f^\dagger}

\def\ket{\rangle}

\def\bran{\langle n|}

\def\phiAn{\phi_{An}^{\hfill}}

\def\bavs3{BaVS$_3$}

\def\t2g{$t_{2g}$}
\def\eeg{$e_{g}$}
\def\a1g{$a_{1g}$}

\begin{document}

\title{Considerable non-local electronic correlations in strongly doped 
Na$_x$CoO$_2$}
\author{Christoph Piefke}
\affiliation{I. Institut f{\"u}r Theoretische Physik,
Universit{\"a}t Hamburg, D-20355 Hamburg, Germany}
\author{Lewin Boehnke}
\affiliation{I. Institut f{\"u}r Theoretische Physik,
Universit{\"a}t Hamburg, D-20355 Hamburg, Germany}
\author{Antoine Georges}
\affiliation{Centre de Physique Th\'eorique, \'Ecole Polytechnique, CNRS, 91128
Palaiseau Cedex, France}
\author{Frank Lechermann}
\affiliation{I. Institut f{\"u}r Theoretische Physik, 
Universit{\"a}t Hamburg, D-20355 Hamburg, Germany}

\begin{abstract}
The puzzling electronic correlation effects in the sodium cobaltate system are
studied by means of the combination of density functional theory and the
rotationally invariant slave boson (RISB) method in a cellular-cluster approach. 
Realistic non-local correlations are hence described in the short-range regime 
for finite Coulomb interactions on the underlying frustrated triangular lattice. 
A local Hubbard $U$ is sufficient to model the gross in-plane magnetic behavior with 
doping $x$, namely antiferromagnetic correlations at intermediate doping 
and the onset of ferromagnetic order above $x$$>$3/4 with a mixed phase for 
0.62$<$$x$$<$3/4. Important insight is thereby provided by the occupations of local 
cluster multiplets retrieved from the RISB framework. The extended modeling of the 
$x$$\ge$$2/3$ doping regime with an additional inter-site Coulomb repulsion 
$V$ on an experimentally verified effective kagom\'e lattice allows to account for
relevant charge-ordering physics. Therewith a fluctuating charge-density-wave state 
with small quasiparticle weight and a maximum in-plane magnetic susceptibility may 
be identified at $x$$\sim$3/4, just where the magnetic ordering sets in.
\end{abstract}

\maketitle

\section{Introduction}
The investigation of frustrated spin systems plays a vital role in condensed 
matter physics due to its immediate importance for the understanding of
competing interactions~\cite{die05}. However many realistic frustrated systems are 
insulating and may not be metalized via doping. 
Hence studying frustration effects within the metallic regime and possibly tuning
interactions by additional doping can often only be realized in a pure model
context. Nonetheless, there is growing awareness that such effects are also 
important in correlated metals~\cite{lac10}. For instance it is believed by many 
researchers that for the recently discovered high-temperature superconducting iron 
pnictides~\cite{kam08}, spin and/or orbital frustration is a relevant 
feature~\cite{si08,kru09}. 

Concerning the doping possibility of a strongly correlated metal with 
manifest geometrical frustrations, sodium cobaltate Na$_x$CoO$_2$ 
serves as an unique realistic test case. This system consists of stacked 
triangular CoO$_2$ layers, held together by Na ions inbetween. Nominally, the 
oxidation state of cobalt ranges between Co$^{4+}$($3d^5$) and Co$^{3+}$($3d^6$), 
depending on the doping $x$. Because of the strong ($t_{2g}$,\eeg) crystal-field (CF) 
splitting a low-spin state is believed to be realized for the Co ion, where $x$ 
controls the final filling of the rather localized $t_{2g}$ manifold. Due to the 
additional trigonal $t_{2g}$-internal $a_{1g}$-$e_g'$splitting, $x$ thus formally mediates 
in a simplified view between a remaining half-filled $a_{1g}$ orbital ($x$=0) and the 
band-insulating regime ($x$=1). In this respect however the trigonal CF splitting 
appears to be a crucial parameter. Calculations based on the local density 
approximation (LDA) yield a value $\Delta(a_{1g}-e_g')$$\sim$$-0.1$ eV for $x$=0.5,
decreasing in magnitude away from this doping level~\cite{lecproc,pil08}. Yet the 
measured Fermi surface (FS) shows only a single distinct hole-like hexagonal 
sheet, whereas LDA predicts additional $e_g'$ hole pockets. Further investigations
have underlined the subtleties associated with this CF splitting for the
cobaltate system~\cite{mar07,wan08,lie08,bou09}. Note that a single hole-like FS 
sheet would transfer into a low-energy single-band modeling with a negative
nearest-neighbor (NN) hopping $t$. Already on this level, an interesting interplay
between the various degrees of freedom (spin, orbital and charge) is expected, 
where the geometrical frustration together with a large ratio $U/W$ of the 
Hubbard $U$ and the bandwidth $W$ should result in a complex behavior with $x$.

Experimental investigations have indeed revealed a very rich phase diagram, 
incorporating various rather different many-body phenomena and phases. A 
superconducting dome ($T_{\rm c}$$\sim$4.5 K) appears close to $x$=1/3 when 
intercalation with H$_2$O is allowed for~\cite{tak03}, while a metal-insulator 
transition (MIT) is found~\cite{foo04} for $x$=0.5 below $T_{\rm MIT}$$\sim$51K. 
Above $x$=0.5 a disproportionation of charge between the Co sites takes 
place~\cite{muk05,lan08} and a regime of large thermopower is 
detected~\cite{wan03} for 0.71$<$$x$$<$0.84. In view of the possible frustration
effects, the varying magnetic behavior is a central concern. For low doping in the 
range $x$$<$0.5 an overall Pauli-like 
susceptibility is found~\cite{foo04} with however hints for two-dimensional (2D) 
antiferromagnetic (AFM) correlations~\cite{fuj04,yok05,lan08}. Such AFM
correlations are identified to increase towards $x$=0 for the structurally
similar Li$_x$CoO$_2$ system~\cite{kaw09}. Yet isolated CoO$_2$, i.e., $x$=0,  
appears to be metallic, whereby the degree of electronic correlations is still a 
matter of debate~\cite{vau07,kaw09}. The case $x$=0.5 seems exceptional, with 
in-plane AFM order~\cite{men05} in the insulating regime. For $x$$>$0.5 the system 
displays increasing magnetical response with a spin-fluctuation regime in the
range 0.6$<$$x$$<$3/4, including the evolution to 
Curie-Weiss behavior~\cite{foo04} for 0.6$<$$x$$<$0.67, and the eventual 
onset of in-plane ferromagnetic (FM) 
order~\cite{sug03,mot03,boo04,iha04,sak04,bay05,hel06,shu07,lan08,schu08}. 
The magnetic structure in the doping range 3/4$<$$x$$<$0.9 with ordering 
temperatures $T_{\rm N}$$~\sim$19-27K~\cite{sug03,mot03,boo04,bay05,men05} is of 
$A$-type, i.e., the FM CoO$_2$ layers are coupled antiferromagnetically. 

The fact that not the low-doping regime with the nominal Co$^{4+}$ ($S$=1/2) 
state (close to $x$=0) but the high-doping region with the nominal Co$^{3+}$ 
($S$=0) state (close to $x$=1) is the magnetically active part in the phase
diagram has motivated various theoretical efforts. Early LSDA 
computations~\cite{sin00,sin03} predicted an itinerant FM solution for all doping 
levels between 0.3$<$$x$$<$0.7, clearly in contradiction with experimental 
findings. In more recent calculations Johannes {\sl et al.}~\cite{joh05}
were able to show that an $A$-type AFM structure is slightly favored over the FM 
state for a simple approximant to the realistic cobaltate crystal structure at 
$x$=2/3. Generally the different ordered magnetic LSDA states are indeed rather 
close in energy~\cite{sin00}. 

Aside from the already named indications, the relevance of correlation effects
beyond LDA due to the partially filled Co($3d$) shell has been motivated by 
several experiments, e.g., from optics~\cite{wan04}, 
photoemission~\cite{val02,has04,yan07,gec07} and transport~\cite{foo04} 
measurements. In view of the cobaltate physics, Merino {\sl et al.}~\cite{mer06} 
discussed in detail the electron-doped single-band Hubbard model with NN hopping 
on the triangular lattice within single-site dynamical mean-field theory (DMFT).
There it was concluded that for $t$$>$0 the paramagnetic (PM) 
Curie-Weiss metallic phase has an instability towards FM order for large enough 
$U/W$. However such an transition is missing for the case $t$$<$0, where 
moreover the PM metallic phase displays Pauli-like magnetic response. In 
Na$_x$CoO$_2$ the dominant NN hopping appears indeed to be negative, but
relevant higher hopping amplitudes (at least to 3rd order) are necessary to 
bring the tight-binding (TB) modeling in good accordance with the LDA 
dispersion~\cite{ros03,joh04}. These higher contributions introduce a flatness
to the low-energy band, giving rise to substantial additional structure in the 
density of states (DOS) compared to the NN-TB DOS~\cite{lee04,lecproc}. 
This allows for a more interesting magnetic behavior with electron doping in 
contrast to the simplified NN-TB ($t$$<$0) model. In fact the
study of Korshunov {\sl et al.}~\cite{kor07} based on the combination of the 
realistic three-band ($a_{1g}$, $e_g'$) dispersion with a Hubbard on-site 
interaction resulted in (${\bf q}$=0)-peaks in the magnetic susceptibility at 
around $x_p$$\sim$0.56 ($U$=0~eV), $x_p$$\sim$0.6 (infinite-$U$ Gutzwiller) and 
$x_p$$\sim$0.68 (infinite-$U$ Hubbard I). Gao {\sl et al.}~\cite{gao07} obtained
in similar calculations using a single-band third NN-TB model also within 
infinite-$U$ Gutzwiller a corresponding value $x_p$$\sim$0.67. In a 
charge-selfconsistent doping-dependent LDA+Gutzwiller treatment with three 
correlated ($a_{1g}$, $e_g'$) orbitals the $A$-type AFM order was found to be
stable for $x$$>$0.6 ($U$=3 eV) and $x$$>$0.5 ($U$=5 eV) by 
Wang {\sl et al.}~\cite{wan08}, but with additional findings of FM order for 
$x$$<$0.3. Note that all these correlated studies neglect non-local effects of 
strong correlation, i.e., the self-energy is taken to be local and, importantly, 
a single correlated Co site in the respective planar part of the unit cell is 
assumed. Moreover the doping dependence of the system is simulated via simple 
band filling or the virtual crystal approximation (VCA)~\cite{nor31}. In a 
single-site DMFT study by Marianetti and Kotliar the importance of the Na 
positions for the strong correlation effects was actually 
demonstrated~\cite{markot07}.

Interestingly, another cobaltate modeling~\cite{hae06} applying the 
$t$-$J$ model with only NN-TB ($t$$<$0) hopping (and simple doping description), 
solved via exact diagonalization on clusters as big as 18 sites (with periodic 
boundary conditions) yield Curie-Weiss behavior for $x$$<$3/4 in contrast to 
the single-site DMFT Hubbard-model studies~\cite{mer06}. In addition recent
experimental findings~\cite{all09} show evidence for a kagom{\'e}-lattice imprint 
due to intricate Na ordering at $x$=2/3. These are only two hints towards the need
for a reinforced description of non-local correlations in the cobaltate theory, 
strengthening also the original aspect of the relevance of the geometric 
frustrations. Both advanced aspects, i.e., strong non-local correlations {\sl and} 
a more elaborate doping dependence within a realistic framework were combined in 
Ref.~\onlinecite{lec09}. There the realistic LDA dispersions for a three-site 
Na$_x$CoO$_2$ in-plane cluster were supplemented with a finite Hubbard interaction 
in a cellular-cluster application of the rotationally invariant slave boson (RISB) 
method~\cite{li89,lec07}. Note that there in the construction of the 
doping-dependent LDA Hamiltonian the respective crystal fields within the 
triangular cluster (modified by close-by Na ions) were not site-averaged as in
the VCA approaches.

In the present work we want to extent the considerations of 
Ref.~\onlinecite{lec09} by a detailed investigation of the cluster multiplets 
in the realistic three-site cluster approach to Na$_x$CoO$_2$. Galanakis 
{\sl et al.}~\cite{gal09} performed a similar study for the doped Hubbard model
on the triangular lattice using the non-crossing approximation (NCA) to cellular
DMFT, but without incoporating realistic band dispersions and without elucidating 
magnetism.

\section{Theoretical Approach}
\subsection{Realistic modeling\label{sec:realmod}}
We describe the sodium cobaltate system from a strongly correlated perspective 
within a realistic framework. This however does not presume that Na$_x$CoO$_2$ may 
be pictured as a doped Mott insulater, a viewpoint often taken for the physics of 
high temperature superconductivity~\cite{leenag06}. Instead here the interplay of 
correlation and frustration effects in an itinerant Fermi-liquid($\mbox{-}$like) state with 
the important proximity to the band-insulating ($x$=1) state appear to be the
main characteristics. The Fermi-liquid character renders the LDA description as a 
reasonable initial starting point. Furthermore the chemical aspects due to the 
intriguing ordering behavior of the sodium ions across the $x$-$T$ phase
diagram brings additional importance to a careful investigation of the realistic
electronic structure with doping. Still because of the narrow bandwidth of the
low-energy sector, even moderate values of the local Coulomb repulsion
on a given Co site puts one in the regime $U/W$$\ge$1. Thus neglecting this fact
may not be adequate on the way to a complete theory of this materials system.

It is aimed for a non-local approach in order to allow for inter-site self-energy
effects on the triangular lattice. To this, as a first approximation, the minimal 
three-site cluster with on-site Hubbard interactions is identified as the 
canonical model. For the lattice description a cellular-cluster methodology shall 
serve as method of choice to combine the NN inter-site correlations with the 
overall itinerant behavior. This extended approach beyond the single-site 
description is supplemented with realistic Kohn-Sham (KS) dispersions from LDA in a 
Wannier-like basis utilizing the maximally-localized 
technique~\cite{mar97,sou01}. The underlying band-structure calculations are 
performed using a mixed-basis pseudopotential implementation~\cite{mbpp_code}. 
This method~\cite{lou79} utilizes norm-conserving pseudopotentials with a 
combined basis set of plane waves and non-overlapping localized functions. 
\begin{figure}[t]
\centering
\includegraphics*[width=8cm]{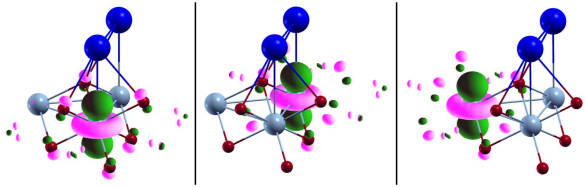}%
\caption{(Color online) $a_{1g}$-like Wannier-like functions
in the Na$_x$CoO$_2$ super cell for $x$=2/3 . Large blue (dark) balls
denote Na ions, light blue (light gray) balls Co ions, small red (gray) balls mark O ions.
\label{real-coba}}
\end{figure}

In the construction of the realistic model the question concerning the 
correlated subspace, i.e., the nature and number of included correlated orbitals, 
arises. In principle the full ($a_{1g}$, $e_g'$) $t_{2g}$ manifold appears as the 
suitable minimal choice. Yet treating a metallic state by a three-site cluster 
with altogether nine interacting orbitals and reduced symmetries in a large doping 
range is numerically very heavy for any well-suited many-body technique. But note 
that the experimental FS studies reveal only a single $a_{1g}$-like band at the Fermi 
level and also the basic LDA studies predict the $e_g'$-like bands close to 
complete filling. 
In addition, in this work we are mainly interested in the magnetic behavior at 
larger doping where in any case these $e_g'$-like bands sink even deeper in 
energy. Hence we reduce the correlated subspace to three effective $a_{1g}$
orbitals per Co cluster, to be determined by the Wannier-like construction and 
therefore including hybridization effects with the remaining orbitals.
This approach seems also suitable from a local-correlation point of view
since the (nearly) filled $e_g'$ orbitals are in a singlet state (low-spin Co) 
and thus do not give rise to additional spin scattering.
In other words, double-exchange processes are expected to be irrelevant. At small
doping however the inclusion of the $e_g'$ orbitals seems necessary because of
an apparent increased relevance of the multi-orbital character. For the 
interesting magnetic properties at larger doping one expects the non-local 
correlations to dominate any intrasite multi-orbital correlations in this
 specific filling scenario.

A meaningful and numerically efficient characterization of the complex 
doping behavior may be build on LDA calculations at selected doping levels. 
A supercell involving the NN Co triangle of a given CoO$_2$ layer serves as the 
base structure. By decorating this triangle with Na ions {\sl above} the Co sites, 
i.e., usually called Na(1) position (see Fig.~\ref{real-coba}), we may represent 
the dopings $x$=$\{0,1/3,2/3,1\}$. Note that in this simplified ansatz no sodium 
ions above/below oxygen positions, i.e., Na(2) positions, are considered.
Albeit this limits the comparability to the true 
structures~\cite{hua04,all09,hua09} it allows for a discrimination of the Co sites 
with $x$ and shall be sufficient for the {\sl overall} behavior with doping. 
In addition we project from the start the obtained complete KS Wannier Hamiltonian 
to the 2D sector and treat only the in-plane dispersion within a given layer. 
Finally the doping-dependent 3$\times$3 cluster band Hamiltonian is constructed 
through linear interpolation:
\begin{equation}
\bH^{\rm KS}(\bK,x)=x\bH^{\rm KS}_{x_a}(\bK)+(1-x)\bH^{\rm KS}_{x_b}(\bK)
\quad,\label{xham}
\end{equation}
where $x_a$,$x_b$ are the neighboring LDA-treated dopings and $\bK$ denotes the
wave vector in the supercell formalism. Note that while VCA averages 
in {\sl real} space, this kind of interpolation takes place in {\sl reciprocal} 
space. Thereby the local differences between the Co sites remain vital
throughout the doping scan. Note that the 2D-only approach is generalized to the
full 3D case within the effective-kagom{\'e}-lattice modeling discussed in 
section~\ref{coba-ext}.

The complete many-body Hamiltonian for a given $x$ to be solved within the 
cellular-cluster approach then reads as follows
\begin{equation}
H=\sum_{\bK ij\sigma}\varepsilon_{\bK ij}\,
d^\dagger_{\bK i\sigma}d^{\hfill}_{\bK j\sigma}+\sum_{\alpha}H_{\alpha}\quad,
\label{eq:fullham}
\end{equation}
where $i,j$ are sites within the triangular cluster $\alpha$, $\sigma$ is the
spin index and $d^{(\dagger)}$ is the electron annihilation (creation) operator.
The local cluster Hamiltonian $H_{\alpha}$ shall consist of an quadratic
on-site part, the local Hubbard $U$ interaction term and possibly a NN inter-site
Coulomb interaction term with amplitude $V$, i.e.,
\begin{equation}
H_{\alpha}=\sum_{ij \sigma}\varepsilon_{\alpha ij}\,
d^\dagger_{i\sigma}d^{\hfill}_{j\sigma}+
U\sum_{i}n_{i \up}n_{i \don}+
\frac{V}{4}\sum_{i\neq j,\sigma\sigma'}
\hspace*{-0.1cm}n_{i\sigma}n_{j\sigma'}\,\,,
\label{eq:halpha}
\end{equation}
with $n_{i\sigma}$=$d^\dagger_{i\sigma}d^{\hfill}_{i\sigma}$. To include the Coulomb
coupling {\sl between} NN clusters a mean-field (MF) decoupling of the inter-cluster 
interactions is performed, resulting in an embedding, which
reads (dropping here the spin indices for convenience):
\begin{equation}
\sum_{i\neq j}n_{i}n_{j}\approx
\sum_{i\neq j}\left(n_i\langle n_j\rangle+n_j\langle n_i\rangle-
\langle n_i\rangle\langle n_j\rangle\right)\approx
\sum_{i\neq j} n_i\langle n_j\rangle\quad,
\label{eq:mfhalpha_v}
\end{equation}
where $i$ is a site on the examined cluster and $j$ is a site on the NN cluster. In
the last step leading to eq. (\ref{eq:mfhalpha_v}) it is assumed that
the average value for $n_j$ may be inserted. We neglect non-density-density terms 
in this MF decoupling~\cite{wen10} since their effect should be minor. Hence the 
modified version of eq. (\ref{eq:halpha}) reads
\begin{eqnarray}
H_{\alpha}=&&\hspace*{-0.2cm}\sum_{ij \sigma}\varepsilon_{\alpha ij}\,
d^\dagger_{i\sigma}d^{\hfill}_{j\sigma}+
U\sum_{i}n_{i \up}n_{i \don}\nonumber\\
&&\hspace*{-0.2cm}+\frac{V}{4}\sum_{i\neq j,\sigma\sigma'}\hspace*{-0.1cm}
n_{i\sigma}\left(n_{j\sigma'}+\frac{1}{2}\langle n_{j\sigma'}\rangle\right)\quad.
\label{eq:halpha_newv}
\end{eqnarray}
The factor 1/2 in the MF term is needed to avoid the double counting of the 
inter-cluster Coulomb interaction when summing over $\alpha$ in 
eq.~(\ref{eq:fullham}). In the actual calculation the values 
$\langle n_{j\sigma'}\rangle$ are initially set and then determined self-consistently
via an outer loop to the RISB scheme. In a PM homogeneous case, the 
MF $\langle n_{j\sigma'}\rangle$ is independent of $j$ and $\sigma'$ and hence the contribution of that 
term corresponds to a mere chemical-potential shift.
The justification of an (extended)-Hubbard-type Hamiltonian for sodium 
cobaltate may not straightforwardly be verified, because additional interaction
terms could arise in a minimal model. However we believe this approach to be 
reliable for most of the key physics in cobaltates. Explicit exchange terms due the 
fact that we integrated out the oxygen states are expected to be small in size and 
are very difficult to compute for metals in a rigorous manner~\cite{joh05}.

The dispersions are provided by the KS Hamiltonian of eq.~(\ref{xham}) (spanned
on $N_K$ points in reciprocal space) in the following way:
\begin{eqnarray}
\varepsilon_{\bK ij}&=&H^{\rm KS}_{ij}(\bK,x)
-\frac{1}{N_K}\sum_{\bK}H^{\rm KS}_{ij}(\bK,x)\quad,\\
\varepsilon_{\alpha ij}&=&\frac{1}{N_K}\sum_{\bK}H^{\rm KS}_{ij}(\bK,x)\quad.
\label{eq:locham}
\end{eqnarray}
Thus the dispersive part carries no local terms, all of those are transfered
into $H_{\alpha}$. In the solution of the problem one has to be aware of the
breaking of translational invariance in the cellular cluster scheme (for 
reviews of cluster approaches to strongly correlated systems see e.g.
Ref.~[\onlinecite{lic03,bir04,mai05}]). Since no detailed K-dependent properties
are discussed, no periodization of the resulting cluster self-energy is 
performed in the present work.

\subsection{RISB method}

The full Hamiltonian from eq.~(\ref{eq:fullham}) is solved employing the 
rotationally invariant slave boson (RISB) formalism~\cite{li89,lec07} in the 
saddle-point approximation. Within RISB the electron operator 
$d^{\hfill}_{i\sigma}$ is represented as 
$\underline{d}_{i\sigma}$=$\hat{R}[\phi]^{\sigma\sigma'}_{ij}f_{j\sigma'}$, 
where $R$ is a non-diagonal transformation operator that relates the physical 
operator to the quasiparticle (QP) operator $f_{i\sigma}$. The transformation 
$\hat{R}$ is written in terms of the slave bosons $\{\phi_{An}\}$.
In this generalized slave-boson theory $\phi$ carries two indices, namely $A$ 
for the physical-electron state and $n$ for the QP Fock state. It follows that 
the kinetic $K$-dependent part of eq. (\ref{eq:fullham}) is expressed via 
the QP operators with renormalized dispersions. The operator character of
the local cluster part is described solely via the slave bosons~\cite{lec07}.
In order to facilitate this operator decomposition into an QP part and the
interacting part carried by the slave bosons, two constraints have to be
enforced. They shall normalize the total boson weight to unity and ensure
that the QP and boson contents match at every site, respectively:
\begin{eqnarray}
\sum_{An}\phi^\dagger_{An}\phiAn&=&1\quad, \\ 
\sum_{Ann'} \phi^\dagger_{An'}\phiAn\bran\fd_{i\sigma} 
f_{j\sigma'}^{\hfill}|n'\ket&=&f_{i\sigma}^{\dagger}\,f_{j\sigma'}^{\hfill}
\quad.\end{eqnarray}
This selection of the physical states is imposed through a set of Lagrange 
multipliers $\{\lambda\}$. In the mean-field version at saddle-point these
constraints hold on average, with the bosons condensed to $c$
numbers. The solved saddle-point equations for $\{\phi\},\{\lambda\}$ yield a 
free-energy $F$ of the interacting electron system that may be used to investigate 
phase stabilities. Of course, this free energy shall not be confused with the total
free energy of the system that also incoporates further terms, e.g., the nuclei
contributions. However already therefrom, many important insights in the 
possible instabilities can be retrieved.
Note that the K-point integration with a smearing method (Fermi, Gaussian)
introduces a small effective temperature into the system, which however may
not easlily be interpreted as a true physical temperature due to the 
fully-condensed boson treatment in this simplest mean-field approach. 

The RISB physical self-energy of the $d$ electrons at saddle-point is given 
by~\cite{lec07}
\begin{equation}
\mathbf{\Sigma}_d(\omega)=\omega\left(1-\mathbf{Z}^{-1}\right)\,
+[\mathbf{R}^\dagger]^{-1}\mathbf{\Lambda}\mathbf{R}^{-1}-\efat^{(0)}\,\,,
\label{eq:Sigma_physical}
\end{equation}
where ${\bf Z}$=$\bR\bR^{\dagger}$ is the QP-weight matrix and ${\bf\Lambda}$ is
the lagrange-multiplier matrix. The quantity $\efat^{(0)}$ is identical to the 
one-body term in the local Hamiltonian and hence in the present case its elements
are given by eq.~(\ref{eq:locham}). Thus $\Sigma_d$ contains only a 
term linear in frequency and a static part. Though more approximative than
involved methods like, e.g., quantum Monte-Carlo, which may handle the full 
frequency dependence, this is sufficient to describe the QP nature at low-energy.
It also gives an idea of the effect of local excitations within a static
self-energy. In this regard, the slave-boson amplitudes yield direct access to
the occupation of local multiplets in the metallic state. The RISB mean-field
approach is moreover rather efficient with very reliable qualitative (and often 
even good quantitative) results in most cases. Note that though a lattice 
implementation of the this method is used here, the technique may equally well 
be utilized as an impurity solver for single-site DMFT~\cite{fer09}. Due to the 
inherent local nature at saddle-point, this specific DMFT impurity solution is 
then identical to the direct lattice-calculation result.

In the present case the formalism is employed within a
cellular-cluster context, i.e., the off-diagonal components of the
self-energy $\Sigma_d(\omega)$ describe inter-site correlation effects
between the Co atoms in the triangular cluster. To obtain an intuitive
interpretation for the slave-boson amplitudes, we rotate both the
cluster-($A$) and the QP-($n$) index of $\phi_{An}$ into the
eigenbasis $\Gamma$ of the isolated cluster problem. The
transformation matrix $\mathcal{U}(\Gamma)$ is obtained via
simultaneous diagonalization of the commuting set of operators
$\{H_{\alpha}, S^2, S_z, N\}$, where $S$ is the total cluster spin
operator, $S_z$ its z-component and $N$ the particle-number
operator. Since $H_{\alpha}$ is a real, symmetric matrix, one ends up
with a representation
\begin{equation}
\phi^{\hfill}_{\Gamma\Gamma'}=\mathcal{U}^\dagger_{\Gamma A}\phiAn
\mathcal{U}^{\hfill}_{n\Gamma'}
\end{equation}
in which the slave bosons $\phi^{\hfill}_{\Gamma\Gamma'}$ are directly tailored
to the multiplet structure of the local cluster $\alpha$. Note that the given
set of quantum numbers is still not sufficient to classify the cluster 
eigenstates completely. In order to simplify the presentation we limit however 
the discussion to this definition of the $\Gamma$ basis, which
is ample for the physics. In the following, a specific multiplet will be 
denoted $\Gamma_{p,n}^{m}$, where $p$ describes the particle sector, $n$ the 
energy level therein (starting with $n$=0 for the respective ground state)
and $m$ marks the spin state, i.e., singlet 's', doublet 'd', triplet 't' and 
quartet 'q'. Where quantum numbers are stated for a given multiplet, we follow 
standard Dirac notation $|E_{p,n}, S, S_z\rangle$ with $E$ the multiplet energy, 
and $S$,$S_z$ total spin and $z$-component respectively.

\section{NN-TB model on the triangular lattice with local Coulomb interaction}

To give an introduction to our approach and to touch base with other recent
work for the Hubbard model on the triangular lattice, we review here results of 
an NN-TB model with $t$$>$0. Hence $V$ is put to zero in this section.
This model considerations shall set the stage for a subsequent discussion of a more 
realistic model of Na$_x$CoO$_2$ with higher hopping terms and, importantly, a NN 
hopping $t$$<$0. We first investigate the isolated triangular cluster to understand 
the structure of the multiplet states and then relate these findings to the physics 
of the itinerant system. 

\subsection{Cluster states in the local limit\label{sec:nn-atom}}

A detailed study of the Hubbard model on the triangular cluster, comparing the 
cases $t$$\lessgtr$0, has been presented in Ref.~\onlinecite{mer06}. There it 
was shown that the tendency towards ferromagnetism is indeed stronger for 
$t$$>$0 already in the local quantum-chemical problem. Here we only sketch the 
$t$$>$0 case and discuss the $t$$<$0 one in line with the realistic sodium cobaltate 
model in section~\ref{coba-mod}. 

\begin{figure}[t]
\centering
\includegraphics*[height=8cm,angle=-90]{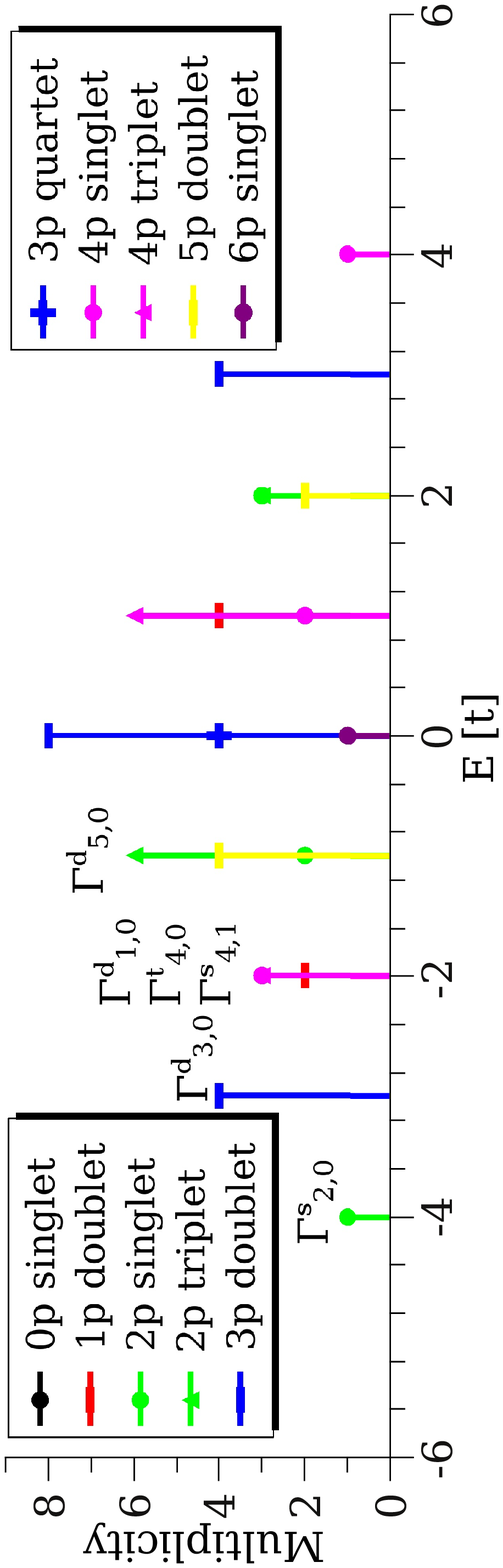}%
\caption{(Color online) Triangular cluster spectrum of $H_{\alpha}$ for $U$=0$t$ and $t$$>$0. 
 The height of each peak shows the respective multiplicity. 
\label{figure:tmSpektrumU=0}}
\includegraphics*[height=8cm]{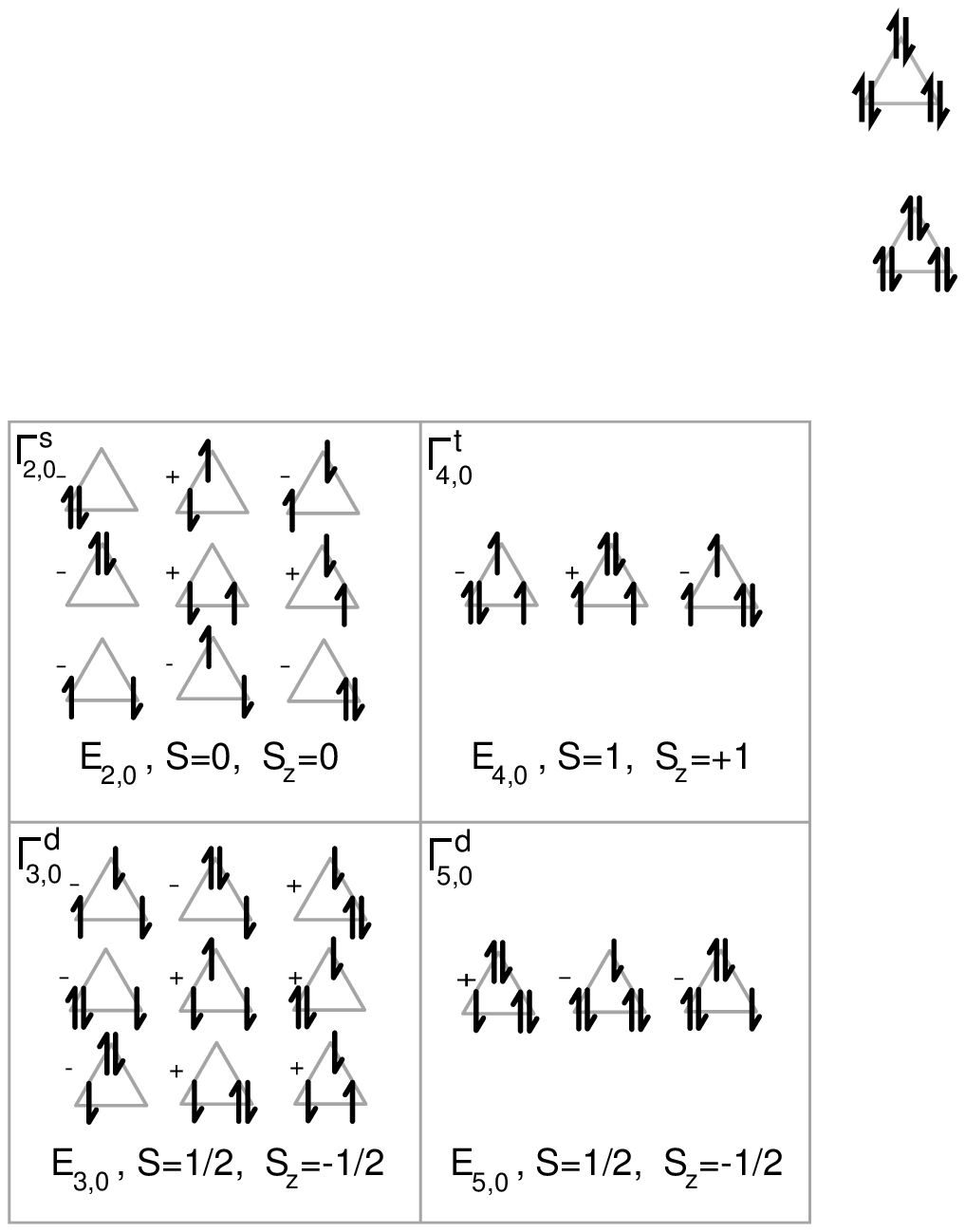}%
\caption{Relevant cluster multiplets for the non-interacting TB model.
  The multiplets are shown as a superposition of Fock states. For
  pictorial simplicity, numerical factors are suppressed. Because the
  factors depend on $U$ and $x$, the Fock states are sorted after the
  absolute value of the factors descending from left to right and top
  to bottom. 
\label{tmStates_non-int}}
\end{figure}

Lets start with the relevant multiplet states on the non-interacting
triangle within the different particle sectors, obtained from a
simultaneous exact diagonalization of $\{H_{\alpha}, S^2, S_z,
N\}$. Note that in the present context $H_{\alpha}$ corresponds to 
eq. (\ref{eq:halpha}) with $\varepsilon_{\alpha ij}$=$-t(1-\delta_{ij})$.
Figure~\ref{figure:tmSpektrumU=0} displays the energy spectrum
of the triangle in the non-interacting case, the decomposition of the
corresponding eigenvectors into Fock states is shown in 
Fig.~\ref{tmStates_non-int} for some relevant multiplets. For pictorial simplicity only the
Fock states participating in a given multiplet are discussed, whereas
the respective expansion coefficients are not shown, but of course are
taken care of exactly within the calculations. In order to give an idea about the
relative size of these coefficients, the pictorial expansion terms in 
Fig.~\ref{tmStates_non-int} (and in the similar following figures) are ordered by 
the decreasing absolute value of the expansion coefficients.

The triangle spectrum of the NN-TB model displays a high degree of symmetry
around zero energy in the non-interacting case. The lowest state in energy
is a two-particle singlet, the highest state is given by a
four-particle singlet. The isolated triangle with NN hopping and
$U$=0$t$ is anti-symmetric under exchange of particles and holes, two doublets from the 
three-particle sector show up at $E$=$\pm3t$. The overall ground state is 
followed by the three-particle doublets, one-particle doublets and
degenerated four-particle singlets and triplets at $E$=$-2t$. As 
pointed out in Ref.~\onlinecite{mer06}, the three different spin-orientations of
the triplet in the four-particle sector are degenerated and the $S_z$ state 
may be changed by spin flips without energy cost. The resulting high susceptibility
for spin polarization in that particle sector is lost for the case $t$$<$0.
\begin{figure}[t]
\centering
\includegraphics*[height=8cm,angle=-90]{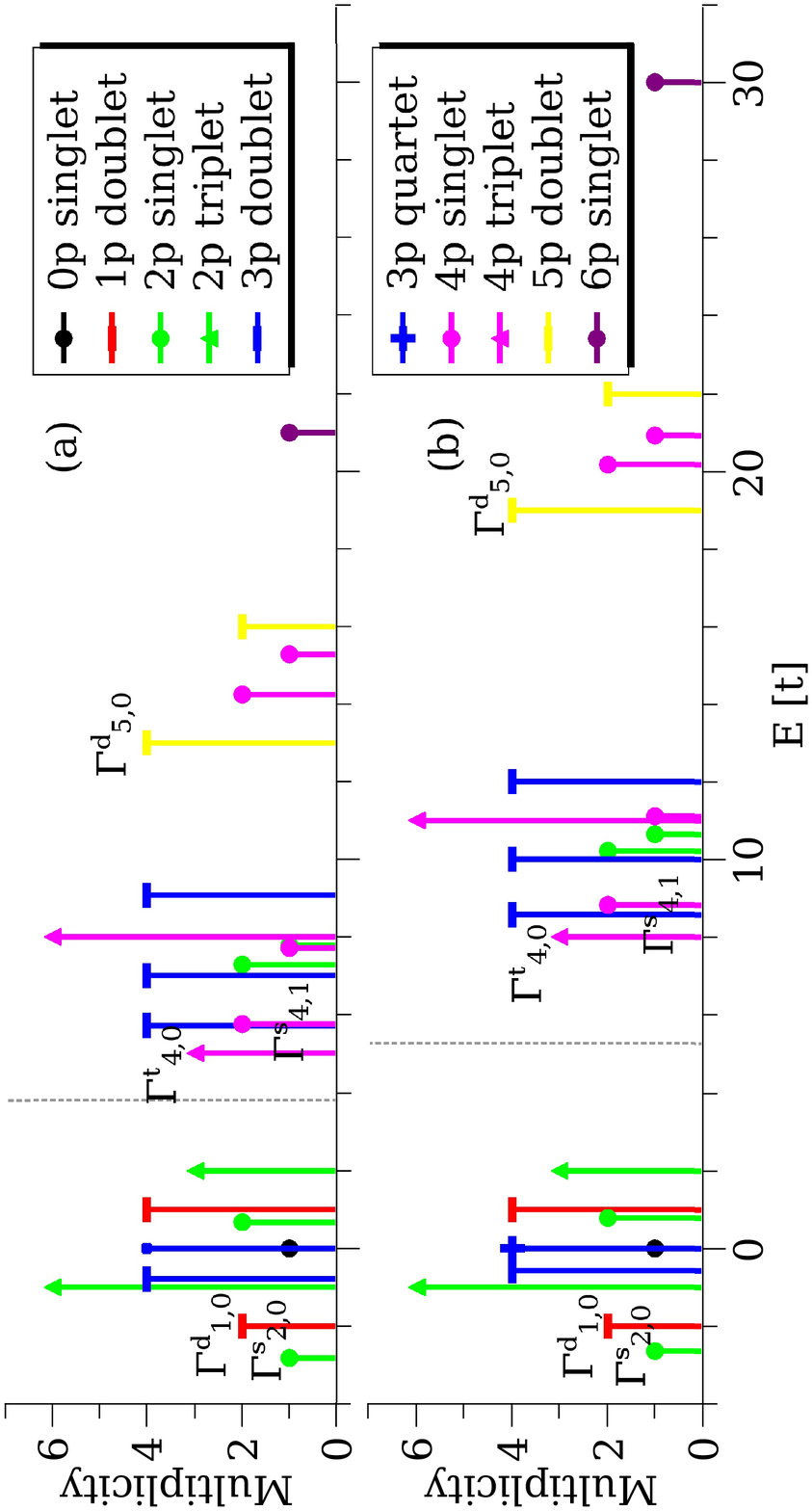}%
\caption{(Color online) Triangular cluster spectrum of $H_{\alpha}$ 
for (a) $U$=$7t$ and (b) $U$=$10t$. Note the level crossing: a 
four-particle sector singlet passes a two-particle sector singlet and a 
four-particle sector triplet at $E$$\approx$$10t$ for $U$=$7t$. No other 
level-crossings where observed for $t$$>$0. The Fermi energy for the half-filled
case of the itinerant problem is denoted by a dashed grey line for comparison.
\label{tmSpektrum}}
\centering
\includegraphics*[height=8cm]{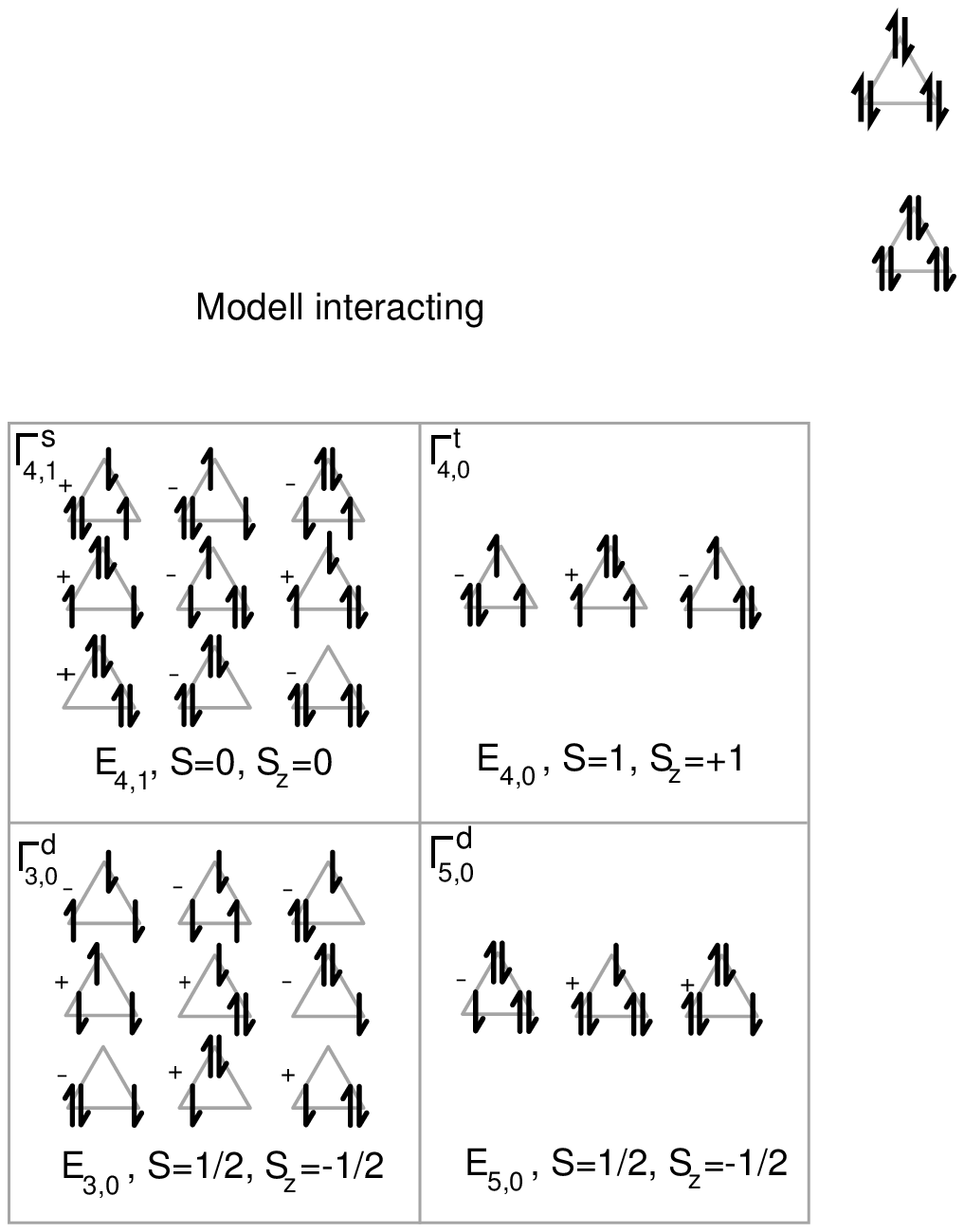}%
\caption{Relevant cluster multiplets for the interacting TB model. 
\label{tmStates_int}}
\end{figure}
Figure~\ref{tmStates_non-int} shows that for $t$$>$0 the structure of the ground 
state in the four-particle sector is rather simple (but becomes more complex in the 
$t$$<$0 case~\cite{mer06}). The simplicity of the dominant doublet 
$\Gamma_{5,0}^{d}$ in the five-particle sector is due to the distribution of a 
single non-local hole, while the three-particle doublet $\Gamma_{3,0}^{d}$ as well 
as the two-particle singlet $\Gamma_{2,0}^{s}$ display a more complex decomposition
into Fock states.

For $t$$>$0 the electrons on the triangle can lower their energy via hopping. The
multiplet $\Gamma_{2,0}^{s}$ contains nine Fock states and each electron in 
those states has two options of hopping, collecting one $-|t|$ every time. From a chemical 
point of view, a bonding state can always be formed. In the case of the three-particle ground state $\Gamma_{3,0}^{d}$, 
the unpaired electron enters an anti-bonding configuration, thus leading to a higher state energy.

Figure~\ref{tmSpektrum} shows the triangular cluster spectrum in the interacting 
regime for $U$=$7t$ and $U$=$10t$, the relevent multiplets in Fock-space decomposition are shown in 
Fig.~\ref{tmStates_int}. For the investigated $U$ values,
the two-particle singlet remains the overall ground state, because it only has to 
pay the $U$ penalty for double occupation three times out of nine. It is energetically followed
by the one-particle doublet which does not gain that much energy from hopping 
processes because of the single-electron occupation. But it now benefits
from being free of $U$ contributions. Hence with increasing $U$ the overall
ground state will eventually evolve towards this one-particle state via a 
``local Mott transition''. Note that the two-particle triplets are also not 
affected by $U$ because they are also free of double occupations. The three-particle
doublet which was following $\Gamma_{2,0}^{s}$ for $U$=0~eV, now has to 
pay the Coulomb penalty six times out of nine. On the other hand the three-particle doublets 
that show the high degree of multiplicity for the non-interacting case at zero
energy split up for finite $U$. It is also interesting to remark that there is a 
swap in energy eigenvalues at $U$$\approx$$9t$, i.e., the energy of a
four-particle singlet increases faster in energy than the other states
in this sector. Besides this one, no further level-crossings were observed with
varying $U$. Again because of the lack of particle-hole symmetry, the two-particle 
sector is favored compared to the four-particle sector even at half filling. 
Note also that the four-particle triplet is lower in energy than the four-particle
singlets. Thus in the atomic limit the triangle is still easy to polarize even for 
$U$$>$0. This holds for the case of three-, four- or five-particle filling, since 
all energetically favorable states from these particle-sectors show a high degree of
multiplicity and have a free spin. While for moderate $U$ the overall ground state 
$\Gamma_{2,0}^{s}$ displays a very complex structure it has no free spin, 
compared to $\Gamma_{3,0}^{d}$.

\subsection{Lattice modeling}

\begin{figure}[b]
\includegraphics*[width=3.25cm]{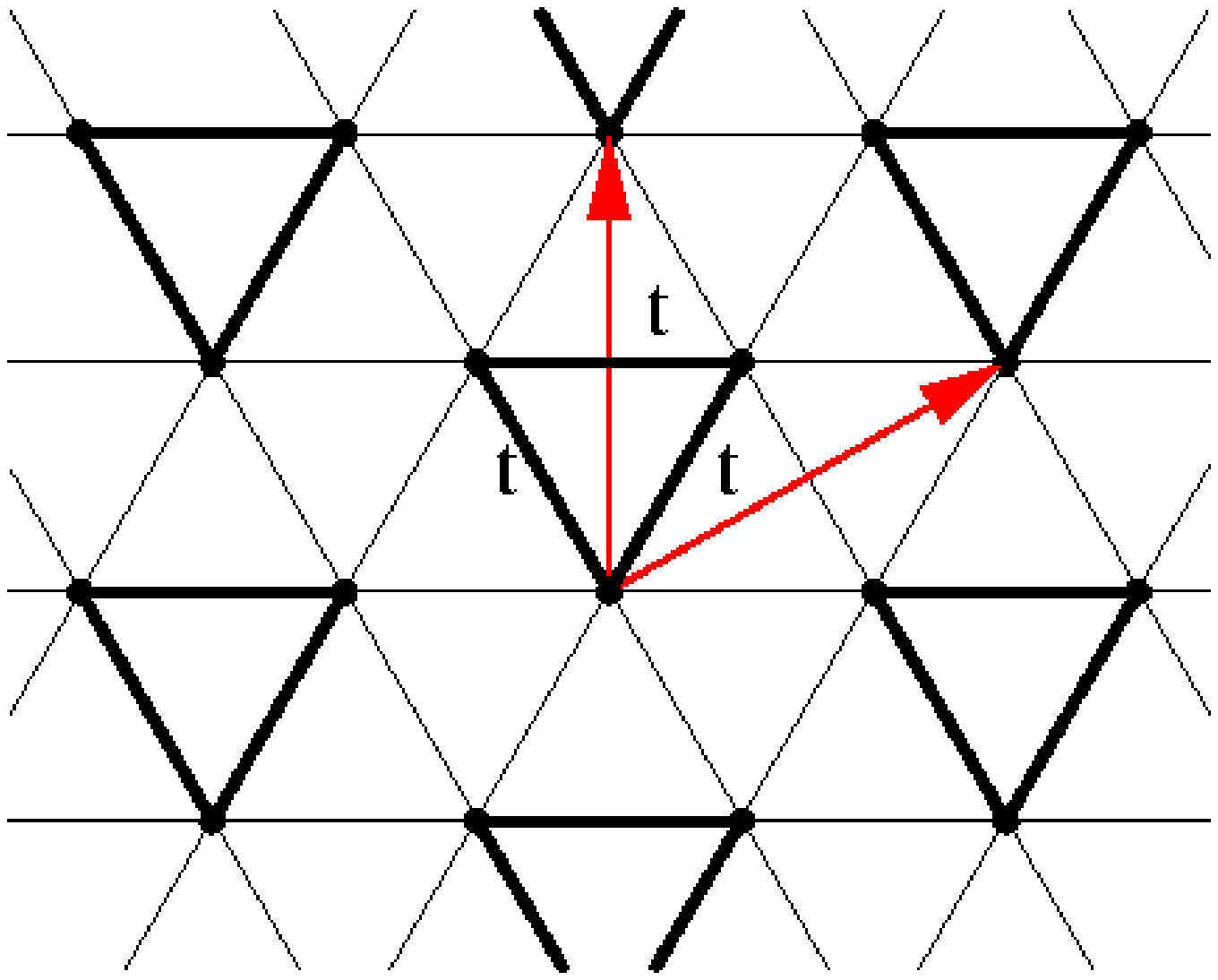}\hspace*{0.2cm}
\includegraphics*[width=4.75cm]{pic/qp-dos.u0.eps}
\caption{(Color online) 3-site cellular cluster approach to the Hubbard model 
on the isotropic triangular lattice for NN hopping $t$. Left: lattice tiling, 
right: density of states with the Fermi energy $E_F$ chosen for half filling .
\label{trilat}}
\end{figure}

In the cellular framework the infinite lattice is build up by finite clusters, here
for the triangular case given by the minimal triangle (see Fig.~\ref{trilat}).
Since we are interested in the canonical 2D lattice, the inter-cluster hopping is 
identical to the intra-cluster one, i.e., the dispersive model is described
by a NN-TB parametrization. In the non-interacting (Fermi gas) case the cellular 
formalism coincides with the standard single-site description, where the bandwidth 
amounts to $W$=$9t$. Because of the absence of particle-hole symmetry the density 
of states (DOS) for smaller filling is strikingly different from the one at larger 
filling (cf.~Fig.~\ref{trilat}). A van-Hove singularity is located at 
$\varepsilon$=$\varepsilon_{\rm max}-t$ ($\varepsilon_{\rm max}$ being the
maximum single-particle energy for the single band), while such a singularity is 
missing at negative single-particle energies $\varepsilon$. Note that hence 
importantly the chemical potential for the half-filled case is {\sl not} placed in 
the middle of the band.

\begin{figure}[t]
\includegraphics*[height=8cm,angle=-90]{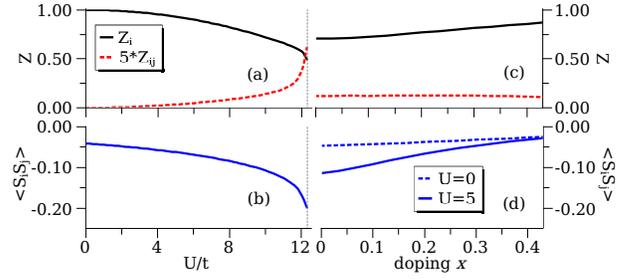}%
\caption{(Color online) Observables for the NN-TB Hubbard model. $U$ dependencies 
for (a) quasiparticle weight and (b) NN $\langle S_i S_j \rangle$ correlation 
functions at half filling. (c,d) same observables, but behavior with doping.
\label{tmZSiSj}}
\end{figure}

For small Hubbard $U$ the interacting model describes a Fermi liquid.
We concentrate in the discussion on the paramagnetic regime of the
half-filled as well as the electron-doped case. The resulting QP weight $Z$ and
inter-site spin correlation $\langle\bf{S}_i\cdot\bf{S}_j\rangle$ on the 
triangular cluster with 
increasing $U$ at half filling are shown in Fig.~\ref{tmZSiSj}(a,b). It is seen that
the diagonal (onsite) QP weight $Z_i$ decreases as $U$ grows, until 
a first-order Mott transition is reached at $U_{c2}$$\approx$$12.2t$=$1.36W$. The 
off-diagonal (inter-site) elements $Z_{ij}$ show opposite behaviour and increase
towards $U_{c2}$, but are much smaller compared to $Z_i$. The literature on the
critical $U$ for the Hubbard model on the triangular lattice, without invoking 
magnetic order, is rather large. The present critical $U$ is in the same range as 
values from exact diagonalization of a finite cluster~\cite{cap01} ($\sim$$12.1t$) 
and cellular DMFT~\cite{kyu07} ($\sim$$10.5t$). Figure~\ref{tmZSiSj}b shows that 
due to the dominant superexchange process at half filling the value for 
$\langle\bf{S}_i\cdot\bf{S}_j\rangle$ is always of AFM character, with a diverging 
slope at $U_c$. 

\begin{figure}[t]
\includegraphics*[height=8cm,angle=-90]{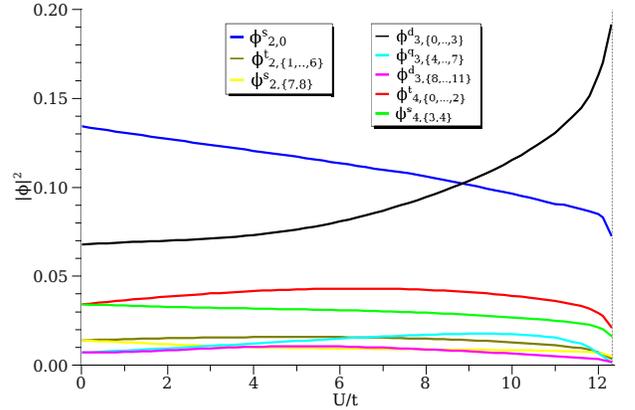}%
\caption{(Color online) Relevant slave-boson amplitudes $\phi$ squared with 
increasing $U$ for the cluster NN-TB model. The amplitudes are labeled with
a single $\Gamma_{p,n}^m$ index (see text). The notation $\{i,\ldots,j\}$ implies 
that the amplitudes for all states from $i$ to $j$ are of equal weight. 
\label{tmUscan}}
\end{figure}

Based on the local-cluster-limit study, the occupation of the
extracted multiplets may now also be investigated in the itinerant problem.
This allows for a discussion of the correlated metallic state for different Hubbard
$U$ and finite doping $x$ from a finite-cluster real-space viewpoint. In this
respect it may be observed in Fig.~\ref{tmUscan} that the cluster multiplet 
occupations are a valuable source of information. When discussing the
condensed slave-boson amplitudes $\phi_{\Gamma\Gamma'}^{\hfill}$ in that 
basis, we focus in the following on the first index $\Gamma$ and sum
 over the second $\Gamma'$ (of course, the full structure was used within 
the calculations). Note that the cluster dispersion on the lattice leads in the 
present case to (weakly) nondiagonal $\phi_{\Gamma\Gamma'}^{\hfill}$ matrices.
For small $U$, the non-degenerate two-particle singlet $\Gamma_{2,0}^s$ has
the highest occupation at half filling. However note that because of degeneracies 
the occupation of several other multiplets is identical, also allowing in principle
for increased fluctuations among them. Thus the total occupation within such a 
class of multiplets, i.e., $\{\Gamma_{4,0}^t,\ldots,\Gamma_{4,2}^t\}$ may be even 
higher.
When normalized to single-multiplet occupations, the three-particle doublet states
$|E_{3,0}^d,\nicefrac12,\pm\nicefrac12\rangle$ dominate for $U$$\gtrsim$$9t$. Close
to the Mott transition the paramagnetic system can be found nearly exclusively
fluctuating between those two doublet states.\\
\begin{figure}[t]
\includegraphics*[height=8cm,angle=-90]{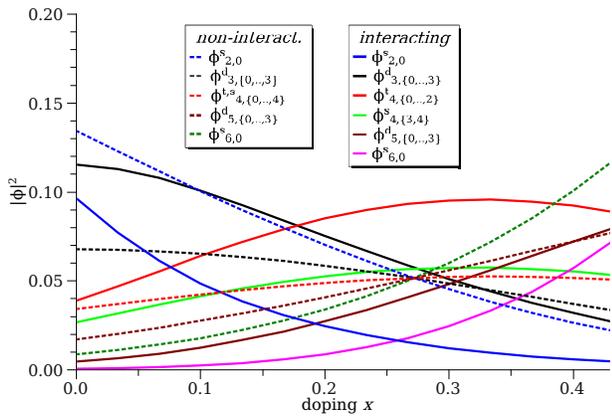}%
\caption{(Color online) Doping-dependent slave-boson amplitudes in the NN-TB model
for $U$=0$t$ and $U$=$10t$.
\label{tmDscan}}
\end{figure}%
Figure~\ref{tmDscan} displays the slave-boson amplitudes with electron doping $x$
up to an investigated regime of $x$=0.425 in the non-interacting case and at fixed 
$U$=$10t$$<$$U_c$. There the $U$=0$t$ case shows a simple crossover from the 
dominant $\Gamma_{2,0}^s$ at half filling to the six-particle singlet 
$\Gamma_{6,0}^s$ at larger filling (with statistical occupation of the remaining 
particle sectors). On the other hand for large $U$ the four-particle triplet 
$\Gamma_{4,\{0,\ldots,2\}}^t$ becomes strongest above $x$$\sim$0.2. Note that the
maximum of $|\phi^t_{4,\{0,\ldots,2\}}|^2$ is reached around $x$=1/3 which is in
the very same range as the onset of ferromagnetism in single-site DMFT for the
given $U$ value~\cite{mer06}. The simple Stoner criterium is known to 
fail in this context~\cite{mer06}, e.g. for $x$$\sim$1/3 the critical $U_c^{\rm FM}$ for 
ferromagnetism on the triangular lattice should be around $4t$. Thus the cluster 
viewpoint provided by the multiplet occupations in the RISB formalism, 
incorporating both, band dispersions and exchange mechanisms beyond 
Hartree-Fock, may serve as a good estimate for the magnetic ordering tendencies on 
the lattice. 
In this respect it is obvious that the large weight of the four-particle triplet 
renders the system rather susceptible to ferromagnetism by aligning the net spin of 
each basic triangle on the lattice. The dominance of the 
$\Gamma_{4,\{0,\ldots,2\}}^t$ state around $x$$\sim$1/3 is also in accordance with 
the recent findings of Galanakis {\sl et al.}~\cite{gal09} in their 
cellular-DMFT study using an NCA impurity solver.

\section{Strongly Doped Sodium Cobaltate With local Coulomb Interaction
\label{coba-mod}}
This section deals with a realistic study of Na$_x$CoO$_2$ over a rather wide
doping range based on projected LDA dispersions and explicit local on-site 
Coulomb interactions. Thus as in the previous section, $V$=0~eV holds here. We will
show that such local Hubbard interactions within a cluster (or multi-site) modeling
are sufficient to deliver a good description of the in-plane magnetic behavior with 
doping. Coming from lower $x$, the PM phase with AFM character as well as the onset 
of FM order at $x$$\sim$3/4 results from this approach. A PM-FM mixed phase 
inbetween these regimes in the area 0.62$<$$x$$<$3/4 is obtained, signaling a 
first-order scenario for this transition and the relevance of disorder/segregation 
effects for a deeper understanding of sodium cobaltate at large $x$.
\begin{table}[t]
\begin{ruledtabular}
\begin{tabular}{r|r|r|r|r|r|r}
$x$ & $\Delta\varepsilon_{\rm Na}$ & $t$ & $t_2$ & $t_3$ & 
$t_4$ & $t_{\perp}$\\ \hline
0          & 0    & -211  & 37    & 33    & -6 & 24 \\
1/3        & -36  & -173  & 34    & 30    & -5 & 10 \\
2/3        & -48  & -134  & 28    & 24    & -4 & 15 \\
1          & 0    & -107  & 20    & 19    & -2 & 20 \\
\end{tabular}
\end{ruledtabular}
\caption{Average hopping integrals for the Na$_x$CoO$_2$ cluster modeling from the 
effective $a_{1g}$ Wannier-like construction shown here up to fourth-nearest 
neighbor (cf. Fig~\ref{trilat}). The on-site level difference
$\Delta\varepsilon_{\rm Na}$ denotes the Na-induced level lowering for Co sites
with sodium on top. Note that in the actual calculations the full 2D hopping 
matrices, i.e., not averaged and not restricted to 4th NN, were employed. 
The inter-layer hopping $t_{\perp}$ is shown for comparison and did not enter the 
RISB calculations discussed in this section. Values are given in meV.
\label{tab:hoppings}}
\end{table}

As described in section \ref{sec:realmod}, the realistic modeling of Na$_x$CoO$_2$
is based on the construction of an effective $a_{1g}$ minimal cluster model.
It is derived from the LDA description of chosen reference configurations 
(serving as simple approximants to the more complex true structures) for the dopings
$x$=$\{0,1/3,2/3,1\}$. The Wannier-like Hamiltonians are projected onto an 
effective single layer, i.e., the RISB method is faced with a 2D
dispersion. The averaged in-plane hopping amplitudes obtained from the respective
Wannier-like construction are given in Tab.~\ref{tab:hoppings}. Compared to the
pure model case in the last sections, the NN $t$ is now negative.
It is moreover evident that the absolute value of $t$ decreases with increasing
doping $x$, in the end loosing about half its magnitude towards $x$=1. This
fact already points to a diminishing superexchange process at larger $x$.
The higher hopping integrals (shown here up to fourth-NN) decay quickly with 
distance and also decrease in magnitude with doping.
Note that the on-site levels are lowered by putting sodium on top.
This can easily be understood by the attractive Coulomb potential introduced by the 
Na$^+$ ions. In Tab.~\ref{tab:hoppings} the NN inter-layer hoppings $t_{\perp}$ are 
also provided.
They show non-montonic behavior with $x$ and an expected slight increase of
the three-dimensional (3D) character through the ratio $t_{\perp}/t$ with doping.
However these and higher inter-layer hoppings are not taken into 
account in the lattice studies presented in section~\ref{sec:coba-metal}.

\subsection{Cluster states in the local limit}

Similar to the strict model case, we begin with the study of the states in
the single-cluster limit. Here the non-interacting part of the Hamiltonian is
given by the $(0,0)$ column of Tab.~\ref{tab:hoppings} (note that of course just
the NN $t$ enters in that limit). Thus not only $t$$<$0 now holds, but also the 
symmetry of the cluster Hamiltonian is lowered at finite doping because of the 
inequivalent Co sites. Our aim is not to discuss the energy spectrum in the
cluster limit at each doping level in full detail. Only the major differences
to the high-symmetry ($t$$>$0) model case treated before shall be elucidated.
\begin{figure}[b]
\centering
\includegraphics*[height=8cm,angle=-90]{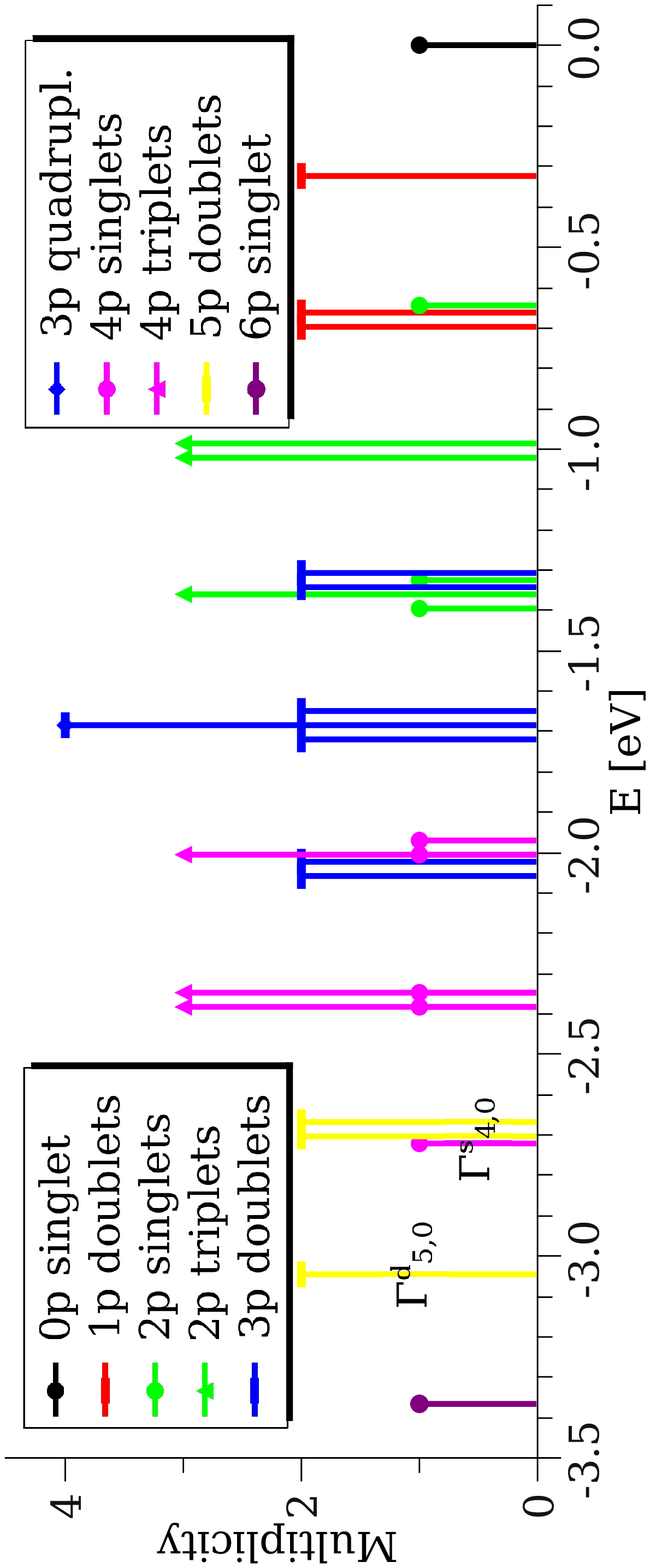}%
\caption{(Color online) Energy spectrum of the cluster limit of the LDA-derived
model Na$_x$CoO$_2$ for $x$=$2/3$ and $U$=0 eV. 
\label{rmSpektrumU=0}}
\includegraphics*[width=8cm]{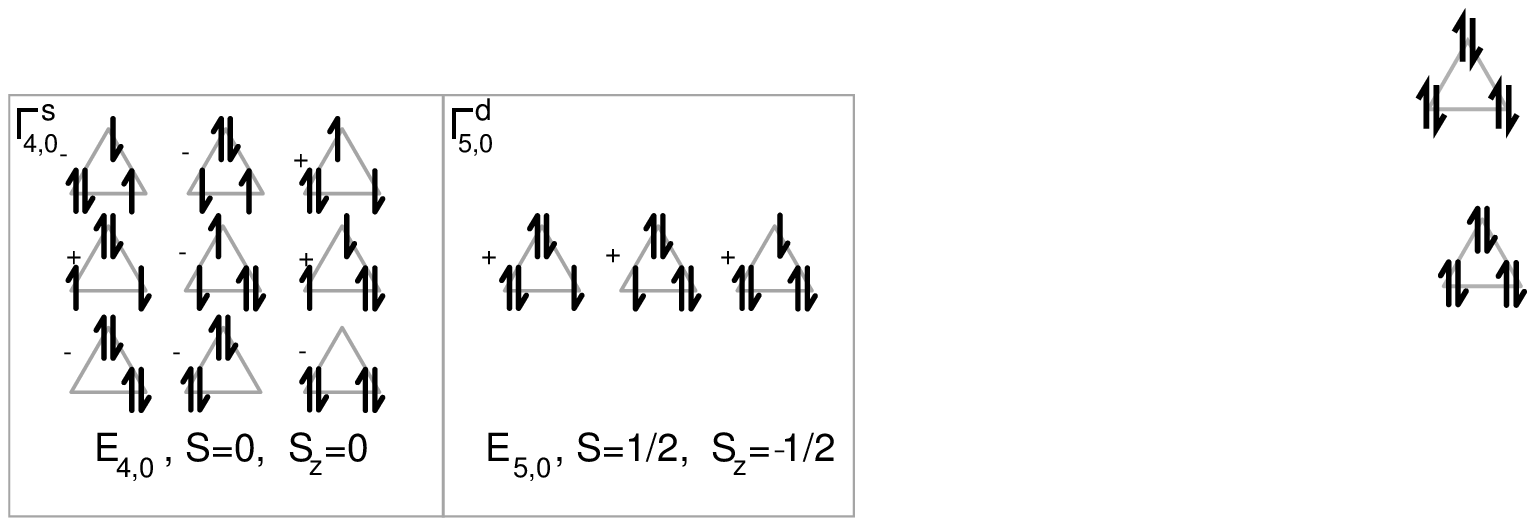}%
\caption{Relevant cluster multiplets for the paramagnetic LDA-derived
     model. 
\label{nacooStates_pm}}
\end{figure}

The spectrum of the cluster eigenstates at $x$=2/3 for $U$=0~eV
is shown in Fig.~\ref{rmSpektrumU=0}. The loss of degeneracy due to the
lower symmetry is obvious. In general, because of the sign change in $t$, the 
spectrum is more or less mirrored compared to the ($t$$>$0) model case. While
for the latter, low-particle sectors where energetically favoured, here  
the six-particle singlet at $\sim$-3.37 eV is the lowest state in energy. Thereafter five-particle doublets
follow in energy, which will turn out to be important in the itinerant lattice 
problem discussed in the next section. As an additional generic difference to
$t$$>$0 the particle sectors tend to group, larger fillings at lower energies due
to the reduced number of possible hopping processes which are now penalized by
energy cost. On the contrary, for the ($t$$>$0) case the particle sectors
are distributed over the whole range of the spectrum for $U$=0~eV
(cf. Fig.~\ref{figure:tmSpektrumU=0}). 

\begin{figure}[t]
\centering
\includegraphics*[height=8cm,angle=-90]{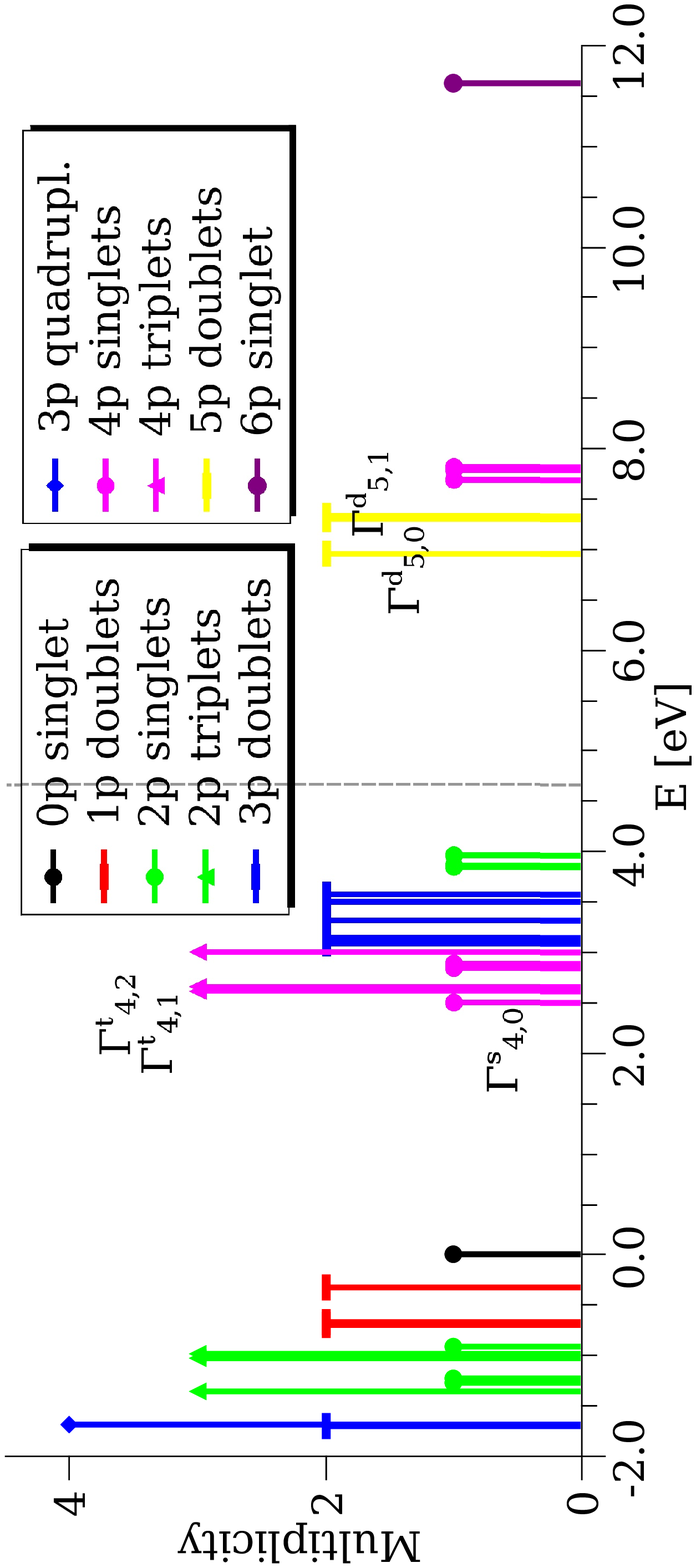}%
\caption{(Color online) Same as Fig.~\ref{rmSpektrumU=0} but for $U$=5 eV.
  \label{rmSpektrum}}
\includegraphics*[height=8cm]{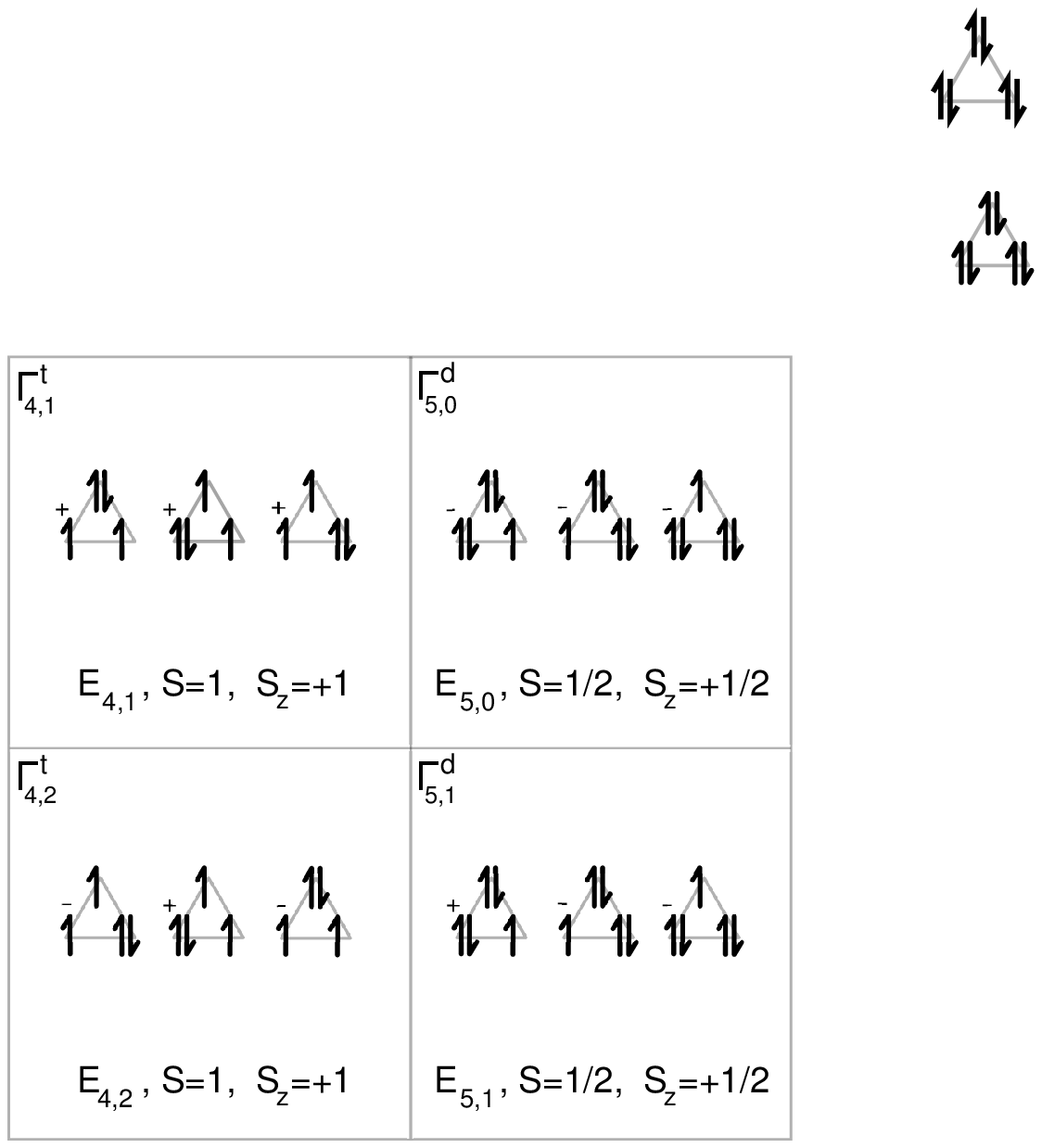}%
\caption{Relevant cluster multiplets for the ferromagnetic LDA-derived
  model. 
\label{nacooStates_fm}}
\end{figure}
Focusing on the four-particle sector which turned out to be important in the 
doped ($t$$>$0) case, it is observable that now a singlet is forming the lowest 
energy state~\cite{mer06}. Remember that the onset of FM order 
for $t$$>$0 from the local viewpoint may be associated with the dominance of the 
ground-state triplet. Moreover the Fock-space decomposition of the four-particle 
ground state, shown in Fig.~\ref{nacooStates_pm}, is far more complex for the 
singlet than for the former triplet (see Fig.~\ref{tmStates_non-int}). The 
ground-state singlet cannot be polarized. To get a polarization, the cluster with 
four electrons would have to be excited, e.g., to the named triplet state. 
But even in the non-interacting case this would cost an energy of about $0.3$eV.

The emerging cluster spectrum in the interacting case ($U$=5~eV) for the Hamiltonian 
associated with the filling $x$=2/3 is shown in Fig.~\ref{rmSpektrum}. Now the 
lowest energy state stems from the three-particle sector and its highly 
degenerated doublets and quadruplets. It is followed by a number of
two-particle states and the overall tendency to group states is enforced. 
Of course by including a Hubbard $U$=5eV the many-particle sectors become 
energetically also costly compared to the non-interacting case.
In the four-particle sector, the singlet is still lowest in energy, but
the triplet draws near, the energy gap becomes smaller by about a
factor of three. That means the interacting Na$_x$CoO$_2$ triangle with
a filling of four electrons is still less polarizable in the
local cluster limit than the triangle with $t$$>$0, but the local Coulomb repulsion
naturally tends to support tendencies towards spin polarization. As shall be seen in the itinerant 
problem, the following three particle doublets and other singlets can be neglected. Next in energy
and easy to polarize due to a single free spin on the cluster is the
degenerated five-particle doublet. As shown in Fig.~\ref{nacooStates_fm} it 
resembles the structure of the easy to polarize four-particle triplet from the ($t$$>$0) case. In other
words, while from this local viewpoint the ($t$$>$0) case exhibits susceptibility
towards FM order at smaller doping associated with the four-particle triplet, the
Na$_x$CoO$_2$ ($t$$<$0) case shows such tendencies only at higher doping connected to the
five-particle doublet.

\subsection{Lattice modeling\label{sec:coba-metal}}

The central aim of this subsection is to investigate the in-plane magnetic 
behavior of sodium cobaltate close to zero temperature with doping $x$. It is known
that eventually the CoO$_2$ planes couple antiferromagnetically (A-type AFM order).
Here we want to address the question why Na$_x$CoO$_2$ displays different
in-plane magnetic correlations with doping and finally FM magnetic order for
3/4$<$$x$$<$0.9. In this respect the mean-field RISB method allows for the
possibility to stabilize paramagnetic as well as symmetry-broken ferromagnetic 
phases for the realistic Hamiltonian (eq.~(\ref{eq:fullham}))~\cite{lec09}. For
instance, a FM phase may be set up via initial spin-different lagrange multipliers
and this lowered symmetry is self-consistently reproduced at saddle-point 
convergence if such a phase is (locally) stable. $U$=5 eV was set 
for the interacting case and we concentrate on dopings $x$$\ge$1/3 where the 
$a_{1g}$-like modeling should be well justified. Note that now of course also the
LDA-derived hoppings beyond NN distance come into play, thus a one-to-one 
comparison with the NN-only model has to be taken with caution.
\begin{figure}[t]
\centering
\includegraphics*[height=8cm,angle=-90]{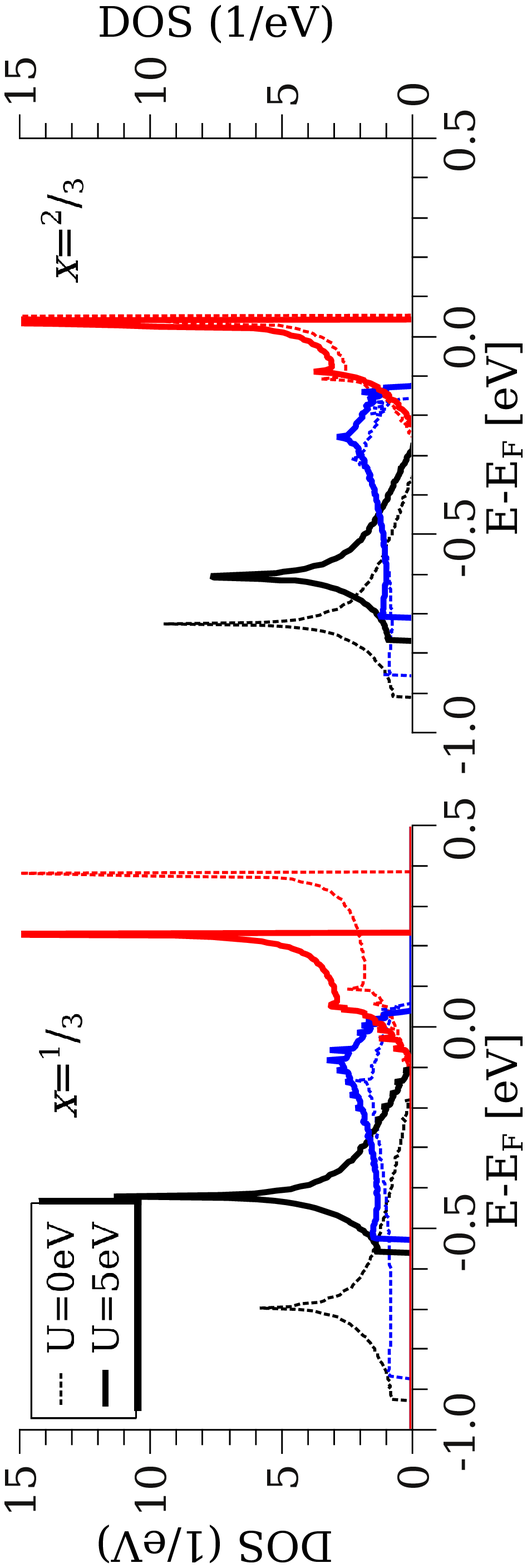}%
\caption{(Color online) QP spectral function of PM Na$_x$CoO$_2$ at $x$=1/3 and
$x$=2/3. The splitting into three parts, respectively, is due to the cluster
description (see text).\label{fig:nacodos}}
\includegraphics*[height=8cm,angle=-90]{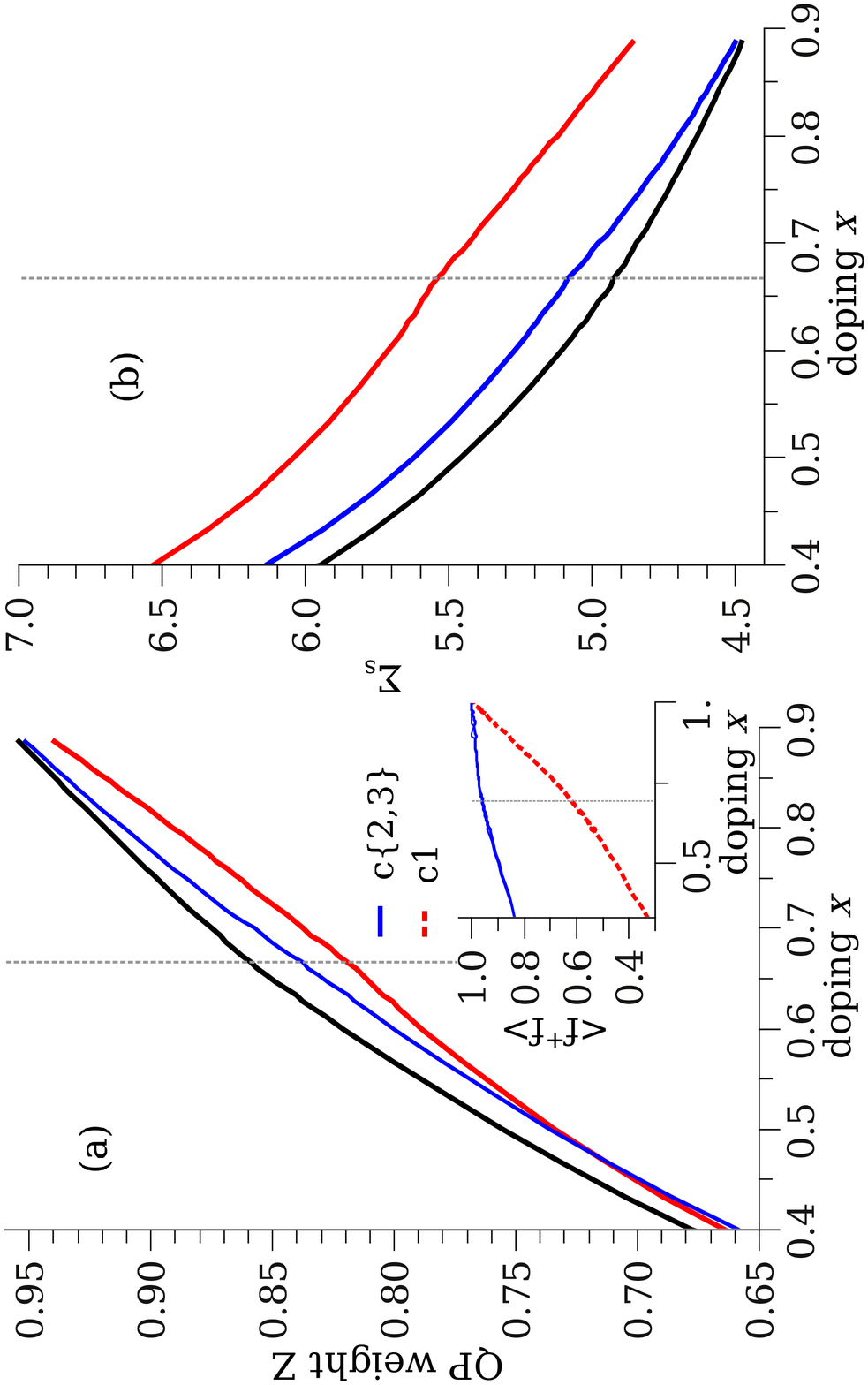}%
\caption{(Color online) Doping-dependent quasiparticle weight (a) and local 
  self-energy (b) for PM Na$_x$CoO$_2$, given in the eigenbasis of the QP 
  operator $\langle f^\dagger f\rangle$, i.e., the orbital density matrix. 
  Close to $x$=2/3 the red/grey 
  orbital, corresponding to the strong filling-dependent state in the inset 
  ($\psi_{\rm c_1}$, also red/grey) shows a stronger renormalisation than the other two (blue/dark). 
  \label{fig:ZSigFdig}}
\end{figure}

The QP spectral function in Fig.~\ref{fig:nacodos} shows the decomposition into
three parts due to the cluster description. The different parts represent the 
K-integrated spectral weight of the resulting supercell bands. They naturally 
add up to the known $a_{1g}$-like cobaltate dispersion~\cite{lee04,lecproc} in the 
non-interacting case. Note that we only consider the 2D
dispersions in K-space. As expected from the doping levels, the overall 
renormalization is weakened at $x$=2/3 compared to the $x$=1/3 case. In the 
higher doping regime the spectral weight strongly increases close to the 
additional van-Hove singularity, which could point to the relevance of Stoner-like
mechanisms for the onset of in-plane FM order.
 
The variation of the QP weight $Z$ and the static self-energy part
$\Sigma_s$ (see eq.~(\ref{eq:Sigma_physical})) with $x$ is plotted in 
Fig.~\ref{fig:ZSigFdig}. We display the quantities in the eigenbasis of the 
cluster orbital density matrix $\langle f^\dagger f\rangle$ of the correlated 
problem in order to make the point that the usual site-basis may not always be the 
most adequate one. Note that resulting eigenstates do {\sl not} directly
correspond to the spectral-weight parts from Fig.~\ref{fig:nacodos}, since the
latter stem from a diagonalization at each K-point. It is seen that two of the
eigenstates of $\langle f^\dagger f\rangle$ , $\psi_{\rm c_{2,3}}$, are highly occupied whereas the 
filling of the remaining effective orbital $\psi_{\rm c_{1}}$ shows strong doping dependence. 
$Z(x)$ and $\Sigma_s(x)$ exhibit the strongest differences among
those states around $x$$\sim$2/3, pointing towards dominant inter-site correlations. Represented in the site basis,
the splitting of the diagonal elements of $Z(x)$ and $\Sigma_s(x)$ is diminished.
However on the present modeling level the discrepancy between the effective 
orbitals, aside from the filling behavior, is not significant. The overall
degree of correlation with $Z$$\sim$0.82-0.86 in this designated doping regime 
is still modest. We will elaborate on this point in section~\ref{coba-ext}.

\begin{figure}[t]
\centering
\includegraphics*[height=8cm,angle=-90]{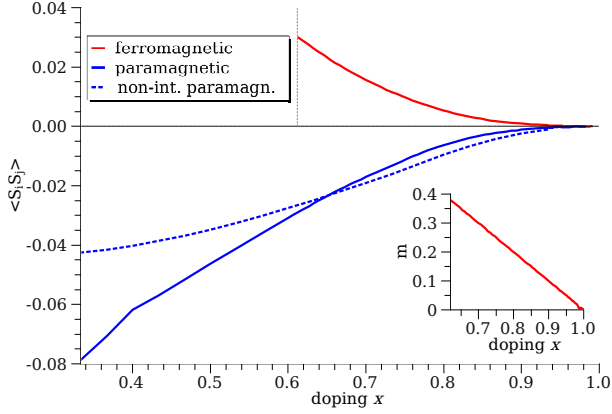}%
\caption{(Color online) Doping-dependent spin correlation function on the 
  triangular cluster for the PM interacting (blue/dark full) and non-interacting 
  (blue/dark dashed) phase as well as the FM phase (red/grey full).
  Inset: averaged cluster magnetization with $x$.
\label{nacoo_sisj}}
\end{figure}

Figure~\ref{nacoo_sisj} shows the pair-averaged spin-correlation functions
$\langle S_i S_j \rangle$ between Co sites $i$,$j$ on the 3-site cobaltate cluster.
Already the non-interacting treatment shows substantial AFM correlations for small 
doping in the PM phase, which are strongly reinforced by the Hubbard $U$. The 
latter is easily understandable from the relevant superexchange mechanism close to 
half filling ($x$=0). But on the other hand interestingly the AFM correlations are 
more significantly suppressed for finite $U$ at large $x$. At $x$$\sim$2/3 there 
is a clear crossover in these short-range correlations visible between interacting
and non-interacting case. Though a generally diminished $\langle S_i S_j \rangle$ 
amplitude at larger doping is reasonable from the sole reduction of the local 
magnetic moments, this crossover remains remarkable. Hence the spin correlations in 
the correlated regime show highly non-trivial behavior with doping. In addition, at 
$x$$\sim$0.61 indeed a long-range ordered FM metal may be stabilized. The metallic 
cluster magnetization of that phase exhibits linearly decreasing behavior with doping,
with subtle convergence properties for $x$$>$0.9 concerning the K-point integration
with smearing methods.

\begin{figure}[t]
\centering
\includegraphics*[height=8cm,angle=-90]{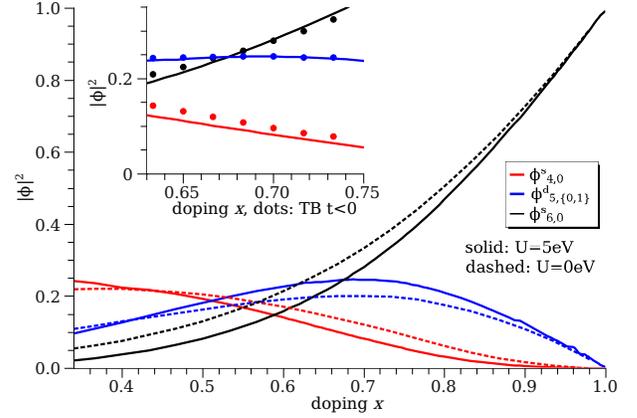}%
\caption{(Color online) Doping-dependent relevant slave-boson amplitudes for
  the interacting ($U$=5 eV) and non-interacting PM phase within
  the LDA-derived Na$_x$CoO$_2$ model. Inset: Comparing interacting cases with
  complete dispersion (full lines) and only NN hopping (dots) around $x$$\sim$0.7.
  \label{nacoo_StatesPm}}
\end{figure}

\begin{figure}[t]
\centering
\includegraphics*[height=8cm,angle=-90]{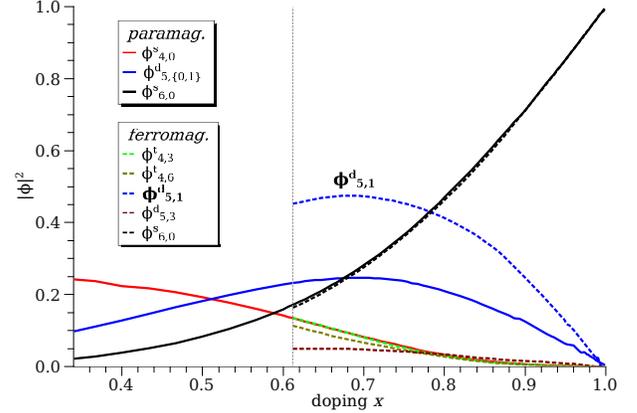}%
\caption{(Color online) As Fig.~\ref{nacoo_StatesPm} but now comparing interacting
PM and FM phases.
\label{nacoo_StatesPmFm}}
\end{figure}

Figures \ref{nacoo_StatesPm} and \ref{nacoo_StatesPmFm} show the
relevant slave-boson amplitudes for the competing phases with doping $x$. In 
Fig.~\ref{nacoo_StatesPm} only the paramagnetic phase is investigated. Since the 
system is in a high-doping regime, multiplets from the four-, five- and 
six-particle sector play the dominant role. Starting below $x$$\sim$0.4, the 
four-particle singlet is strongest, in line with the ongoing NN-AFM correlations
there. At $x$$\sim$0.52 the five-particle doublets however overrun 
the four-particle-singlet weight and reach a maximum at around $x$=0.69. Hence
at $x$$\sim$2/3, the probability to find the system in one of
the two doublet states is about 25$\%$ each. As the doping increases even further
the six-particle singlet, corresponding to completely filled cluster becomes 
of course dominant. As expected, the six-particle singlet is much stronger 
occupied in the non-interacting case for constant $x$. However also
the occupation of the doublet is smaller for $U$=0~eV, while the four-particle 
singlet gains weight in the doping range $x$$>$0.44. This corresponds to an 
enhanced establishment of a net spin on the triangular cluster with sizeable 
Hubbard $U$, which gives the trend towards ferromagnetic behavior on the 
lattice. At small doping below $x$$\sim$0.44, the singlet is favored by 
interaction, underlining the AFM tendencies. It is interesting to note that the
dominant maximum of the five-particle doublet around $x$$\sim$0.7 only little
depends on the influence of the hopping terms beyond NN. Albeit the latter are
responsible for the van-Hove singularity close to the upper band edge for $t$$<$0,
a RISB calculation with only NN $t$ (fixed here to the value $t$=$-135$ meV) 
results mainly in the enhancement of the four-particle singlet weight (see inset
of Fig.~\ref{nacoo_StatesPm}). Thus the more distant hoppings indeed strengthen the
trend towards in-plane ferromagnetism, but it is questionable if this is the only
explanation for the eventual FM order.

 
The occupations of the slave-boson states in the PM and the FM phase are compared
in Fig.~\ref{nacoo_StatesPmFm}. In the ferromagnetic phase, the 
degeneracy of the dominant five-particle doublets is lifted with the favored 
$S_z$=1/2 one showing a prominent maximum at $x$$\sim$0.68. The probability to 
find the system in that state amounts to roughly 50$\%$. Note that the 
four-particle ground-state singlet is not occupied  at all in the FM phase, 
instead triplet states with $S_z=$+1 from that particle sector display some weight.

\begin{figure}[t]
\centering
\includegraphics*[height=7cm,angle=0]{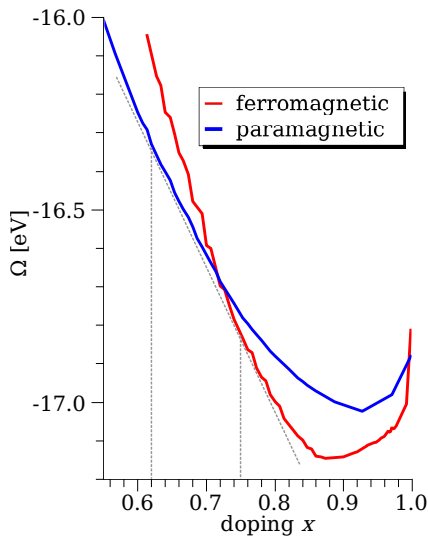}
\caption{(Color online) RISB free energy for the PM (blue/dark) and FM
  (red/grey) phase. There is a region of coexistence ranging from 
  $x_a$$\approx$0.62 to $x_b$$\approx$3/4 obtained from the tie-line 
   construction.
  \label{freeenergies}}
\end{figure}

In order to map out the actual stability ranges of the PM and FM phase,
the doping-dependent free energy has to be studied. Accordingly 
Fig.~\ref{freeenergies} shows the respective curves obtained from the
saddle-point solution of the RISB free-energy 
functional~\footnote{There are some minor differences in the free-energy plot of
Fig.~\ref{freeenergies} compared to the one presented in Ref.~\onlinecite{lec09}.
This is due to the fact that here a smaller gaussian smearing was used in the 
k-point integration and that technical difficulties in the phase stabilization 
for $x$$>$0.9 were solved by improved mixing procedures for the solution of
the saddle-point equations.}. Be aware that our present discussion is limited to 
the 2D limit of Na$_x$CoO$_2$ and that very general stability statements can of 
course only be made for the full 3D case~\cite{lec09}. Note also that a {\sl full} 
FM order of the 3D lattice can truly not be stabilized in the RISB treament of our 
model cobaltate. One may observe in Fig.~\ref{freeenergies} that the FM 
phase is only (meta)stable around $x$$\sim$2/3. Above $x$$\sim$0.7 it becomes 
locally stable, but as revealed by the tie-line construction, globally a mixed 
PM-FM phase is the thermodynamic stable solution in the range 0.62$<$$x$$<$3/4. 
This first-order scenario for the onset of magnetic order is substantiated by the 
experimental findings of A-type AFM ordered phases starting at $x$$\sim$3/4 with 
a small but still sizeable finite magnetic moment (in line with our theoretical 
values). Moreover the peculiar behavior in the given doping range, especially 
around $x$$\sim$0.7, was elucidated in several experimental 
works~\cite{muk04,sak06,bal08,zor10}. The RISB FM free-energy curve displays a
minimum close to $x$$>$0.9. While in experiment the magnetic order is
lost for $x$$>$0.9, our mean-field solution in the 2D limit still allows for 
in-plane FM order up to the band-insulating regime. But the obtained numerical
solutions in that doping regime are rather fragile and temperature effects (somehow
mimiced in the present RISB implementation via the smearing in the K-point 
integration) are very effective in easily destroying a simple thermodynamic phase
behavior. Remember that experimentally close to $x$=1 strong tendencies towards 
phase separation are seen (e.g. Ref.~\onlinecite{lee06} and 
references therein), which may also be detected theoretically~\cite{lec09}.

Since the employed cellular-cluster calculations are numerically rather costly, the question 
arises if the thermodynamic behavior from Fig.~\ref{freeenergies} may also 
be retrieved from a single-site point of view. Therefore we studied two alternative
approaches, namely the multi-single-site solution of our established cobaltate 
model and a true single-site model derived from the LDA band structure of the 
single Co site unit cell of NaCoO$_2$. Note that the former model corresponds to
the neglect of inter-site self-energies, however keeping site-dependent 
self-energies in the cluster unit cell. The calculations revealed that the 
multi-single-site approach yields nearly identical free-energy curves in the 
relevant doping regime, with deviations becoming sizeable only at smaller $x$ 
closer to the half-filled regime. Still, quantitatively the cluster free-energy 
remains always lower by a few ten meV. On the other hand the true single-site model 
showed similar tendencies, i.e., PM phase at intermediate and FM phase at high 
doping, but resulted in rather different phase boundaries not matchable with 
experimental findings. Hence inter-site self-energies seem not essential for the 
{\sl overall integrated thermodynamic} behavior on the present level of modeling 
(but surely are for the detailed electronic signatures). Yet a non-local 
description in the sense of site-dependent self-energies in enlarged supercell 
descriptions is believed to be crucial for the energetics of sodium cobaltate.

\section{Strongly Doped Sodium Cobaltate with extendend Coulomb Interactions
\label{coba-ext}}

So far the modeling was concentrated on the generic doping dependence of the
correlation effects within a single-layer approximation to the range
1/3$\leq$$x$$\leq$1. In this last section we want to elaborate on
the modeling for the doping regime around $x$$\sim$2/3, which appears distinguished
because of its strong thermoelectric repsonse as well as its magnetic behavior
in line with the apparent phase competitions discussed in 
section~\ref{coba-mod}. Hence a more refined structural modeling taking also into 
account the observed charge ordering phenomena for $x$$>$0.5 should shed more light
upon the physics in this region of the phase diagram. Recent experimental findings 
revealed the possibility of a kagom{\'e}-lattice imprint in the CoO$_2$ layers at 
$x$=2/3 due to Na-ordering driven charge order~\cite{all09} and we want to build 
up a more subtle realistic cobaltate model on this proposal. To include the effect
of a charge-ordered background the inter-site Coulomb interaction $V$ is now vital.
As will be shown, therewith one is also in the position to finally describe the
strong mass-renormalization effects in this high-doping regime, which were out of
reach in the local-$U$ only modeling (cf.~Fig.~\ref{fig:ZSigFdig}).

\begin{figure}[t]
\includegraphics*[height=3.5cm]{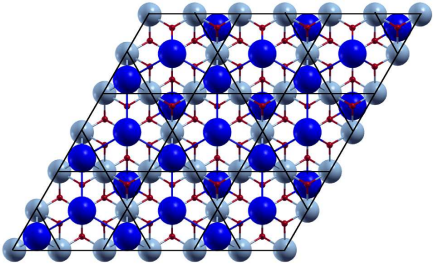}%
\caption{(Color online) Top view on the Na decorated CoO$_2$ layer of the utilized crystal
approximant to the experimentally detected kagom{\'e}-lattice imprint~\cite{all09}. 
Sodium atoms are marked blue/dark, cobalt atoms grey and oxygen atoms red/lightgrey.
The resulting effective kagom{\'e} lattice is sketched via dark lines.  
\label{kaglat}}
\end{figure}

Figure~\ref{kaglat} shows the Na-decorated CoO$_2$ layer of an approximant to the
rather complex true Na$_{\nicefrac{2}{3}}$CoO$_2$ lattice structure detected in 
Ref.~\onlinecite{all09} (incorporating 88 atoms in the unit cell). The former
structure has four in-plane Co sites in the unit cell, where one of the sites
(Co(1)) has a Na ion on top, i.e., in the Na(1) position, and the other sodium 
ions occupy Na(2) positions close to the remaining Co(2) sites. Thereby the 
approximant is a simple approach to mimic the key property of the original 
structure, namely a kagom{\'e}-lattice imprint if one assumes the Co(1) sites to 
be formally in the charge-blocked Co$^{3+}$ state~\cite{all09}. Albeit a simple 
LDA calculation surely does not yield a strong charge-ordered pattern, we proceed in
constructing an effective kagom{\'e} lattice model for 
Na$_{\nicefrac{2}{3}}$CoO$_2$ by projecting the resulting band structure onto
$a_{1g}$-like Wannier functions centered {\sl only} at the Co(2) sites. Since in 
addition we want to incorporate now also the full staggered 3D character we use 
the Co(2) sites in two adjacent CoO$_2$ layers. Hence we end up with a 
six-dimensional model where, respectively, three Wannier functions are again 
located on an in-plane triangle cluster as a basis for an effective kagom{\'e} 
lattice (cf. Fig.~~\ref{kaglat}). Note that this is a simple approximation to an 
actual blocking of the Co(1) sites through projecting out the explicit hoppings 
degrees of freedom associated with orbitals located on those sites. From this 
derivation of the realistic Hamiltonian the NN hopping amounts to $t$=-152 meV, 
i.e. is slightly larger than for the initial triangular lattice model at $x$=2/3 
(cf. Tab.~\ref{tab:hoppings}).
The NN inter-layer hopping amounts now to $t_{\perp}$=9 meV, i.e., is somewhat 
smaller than in the pure triangular case. Thus the 2D character, again loosely
defined via $t_{\perp}/t$, is reinforced for the effective kagom{\'e} lattice.

In the following the emphasis is still on the physics within a single CoO$_2$
layer and in the PM phase. However the description within the full 3D structure, 
allowing here for inter-layer resolution, is capable of revealing not only the 
experimentally-observed bi-layer splitting in the correlated regime. In addition
we now easily may stabilize the A-type AFM order in the RISB calculations. While 
$V$ is varied to look for the influence of the inter-site Coulomb term, the Hubbard
$U$ remains fixed to $U$=5~eV. Since the physics is now studied on an effective 
lattice where the supposingly blocked Co$^{3+}$ states have been projected out, the 
nominal filling on that lattice is of course not identical to the original doping 
level $x$. Its easily shown that the letter value is related to the effective doping 
$x_{\rm eff}$ for the given kagom{\'e} problem through $x_{\rm eff}$=$(4x-1)/3$.
\begin{figure}[t]
\includegraphics*[width=5.5cm,angle=-90]{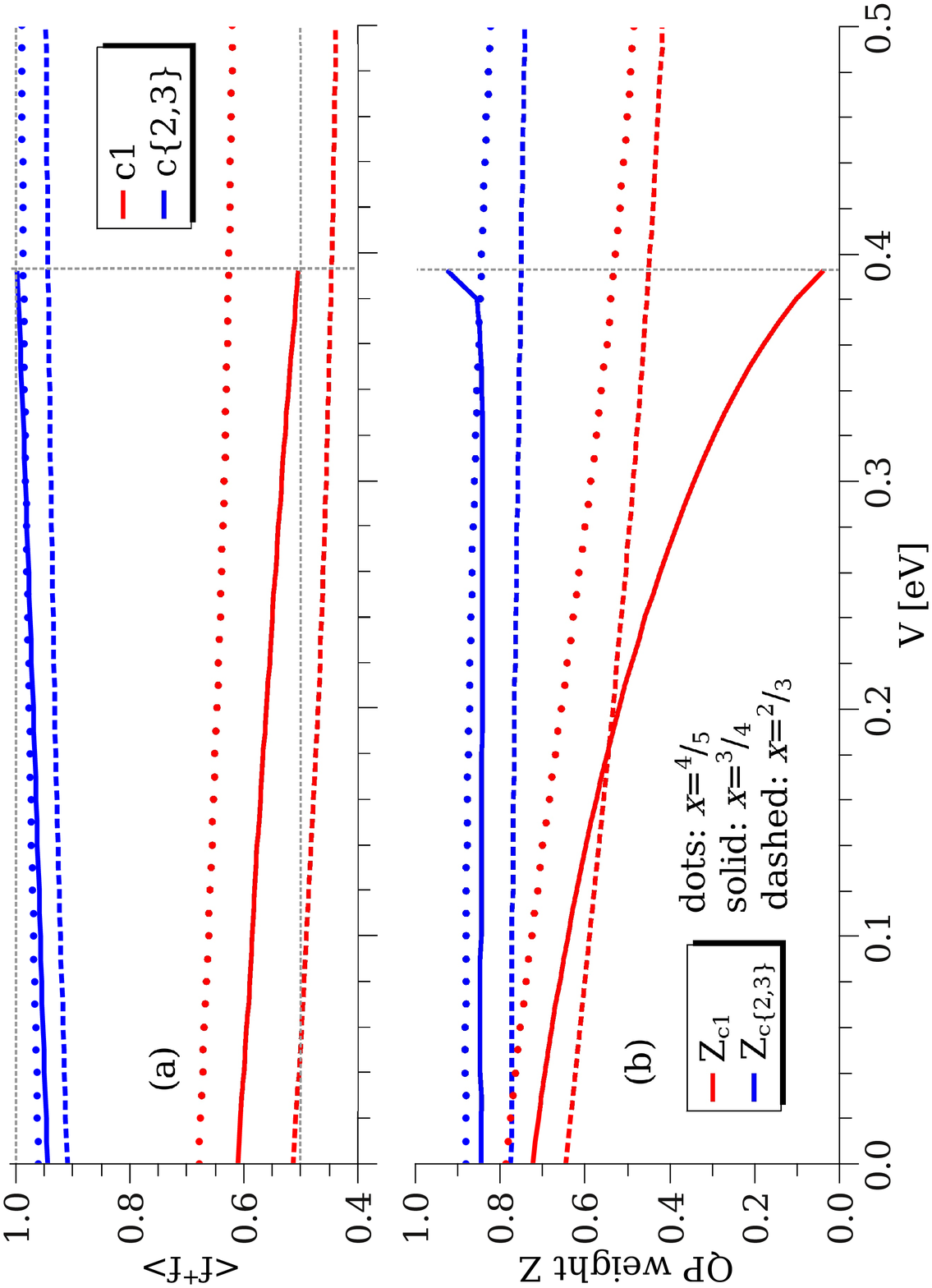}%
\caption{(Color online) Occupations (a) and QP weight (b) for the extendend
cobaltate Hubbard model on the effective kagom{\'e} lattice with increasing 
$V$ in the eigenbasis of the cluster orbital density matrix. 
\label{kagz}}
\includegraphics*[width=5cm,angle=-90]{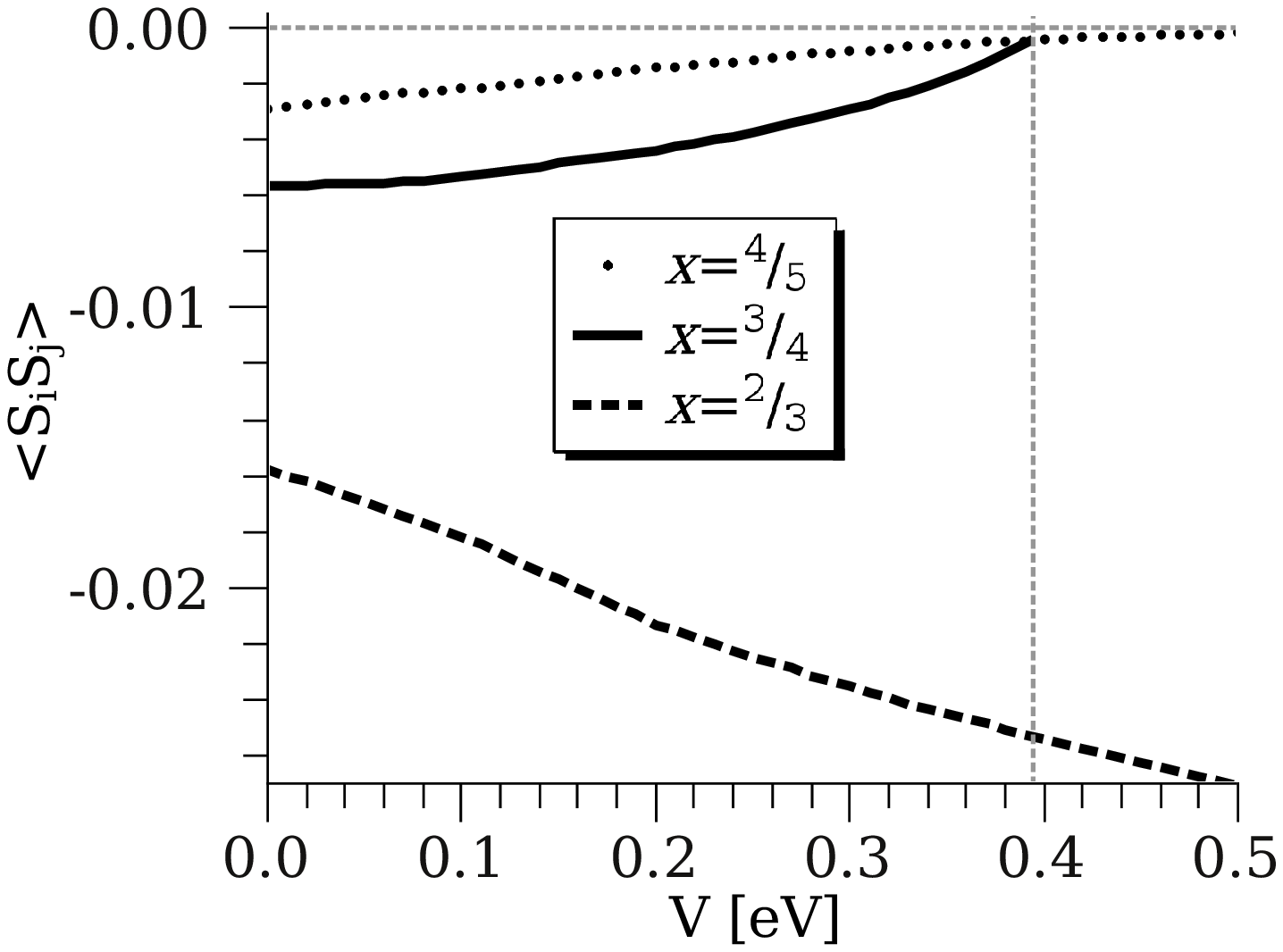}%
\caption{(Color online) Inter-site spin correlations on the triangular cluster of
the  the effective kagom{\'e} lattice for the three different doping levels.
\label{kagsisj}}
\end{figure}

The liberty is taken to allow for a stability of the effective kagom{\'e} 
structure also for dopings somewhat larger than $x$=2/3. Since it is 
known that the charge ordering persists in that region, the use of the
($x$=2/3)-structure with adjusted doping level may serve as a suitable first 
approximation. In Fig.~\ref{kagz} the orbital filling and the QP weight $Z$ of the 
eigenstates of $\langle f^\dagger f\rangle$ for the dopings $x$=2/3,3/4,4/5, i.e.
$x_{\rm eff}$=5/9,2/3,11/15, are displayed. The
transformation to the given eigenbasis becomes now vital since its is easily
observed that the inter-site $V$ has a rather strong influence on the observables.
For $x_{\rm eff}$=2/3 the effective kagom{\'e} lattice is in the ordering 
regime of a charge-density-wave (CDW) instability on a lattice with triangular
coordination~\cite{mot04,has07,bej08,wen10}. In the present study, this 
instability is not driven by a charge localization on the lattice sites $i$, but 
in the eigenstates $c$ of the triangular cluster. The strongly renormalized
state $c_1$ in Fig.~\ref{kagz} corresponds locally to a symmetrical cluster eigenstate
of the form $\psi_{\rm c}$=$A\sum_{i=1}^3\psi_{\rm i}$ (note that $t$$<$0 inverts
the bonding-antibonding energy hierachy), where the amplitude $A$ is
identical for all $a_{1g}$-like site-centered states and independent of $V$.
The remaining two cluster eigenstates are degenerate with non-trivial
linear combination of the site-centered states. Note that this result may serve as
a natural explanation for the findings of small quasiparticle weights in the
high doping regime~\cite{qia06,bro07,nic10}. 

Thus for large $V$ and $x_{\rm eff}$$\sim$2/3 the system is unstable against a
resonating-valence-bond- (RVB) like CDW phase. Because of the
cluster choice within the cellular description, that phase still
breaks translational invariance. At the actual $x$=2/3 level, the
discrimination between the quasiparticle weight of the respective
eigenstates is still visible, yet the strong signature is lost.

The critical value $V_c$$\sim$0.4 eV for the metal-insulator transition at
commensurate filling is perfectly reasonable for the cobaltate
system. Concerning the in-plane spin correlations, Fig.~\ref{kagsisj}
remarkably shows that the inter-site $V$ has a qualitatively different
effect on the various dopings. While $V$ for $x_{\rm eff}$=5/9 ($x$=2/3)
strengthens the AFM tendencies, at $x_{\rm eff}$=2/3 ($x$=3/4) the AFM
character is substantially weakened, with a clear trend towards FM
correlations. For $x_{\rm eff}$=11/15 ($x$=4/5) the NN-AFM correlations are 
already rather weak in the PM phase. 
Interestingly, $x_{\rm eff}$$\sim$2/3 just corresponds
to the actual doping value $x$=3/4 of the full triangular lattice
where the onset of in-plane FM order is found experimentally. In this context
it is essential to note that there are very strong hints that the maximum 
experimental Curie-Weiss susceptibility is indeed not at $x$=2/3 but closer to the
$x$=3/4 doping level~\cite{foo04,bob07}.
Hence besides the general thermodynamic considerations presented in the
former section, the explicit inclusion of realistic charge ordering
effects also points to the onset of in-plane FM order at $x$$\sim$3/4 with a
unique underlying electronic phase.
\begin{figure}[t]
\includegraphics*[width=5.5cm,angle=-90]{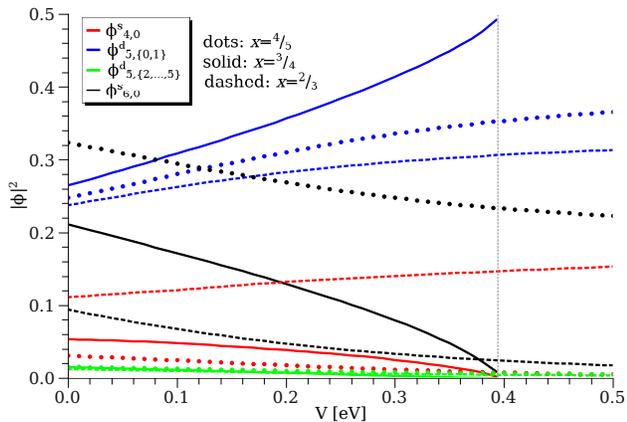}%
\caption{(Color online) Slave-boson amplitudes of selected multiplet states
on the triangular cluster of the effective kagom{\'e} lattice for the three
different doping levels. 
\label{kagmultiplet}}
\end{figure}

Considering once again the slave-boson amplitudes of the local cluster multiplets 
for the investigated doping levels, Fig.~\ref{kagmultiplet} renders it obvious 
that for $x$=3/4 the five-particle doublets are strongly favored by $V$. On the
other hand for $x$=2/3 that multiplet, though still dominating, has to cope with
a still sizeable occupation of the four-particle singlet state. Therefrom,
the enhanced AFM tendencies in the latter instance compared to the $V$=0~eV case
are understandable from the local cluster perspective. This four-particle singlet 
state becomes even more suppressed for $x$=4/5, where experimentally already the
in-plane FM phase is established.

\section{Conclusions}
In this paper we employed the RISB mean-field formalism to approach the 
sodium cobaltate system at larger doping by combining realistic LDA dispersions
with tailored many-body model Hamiltonians in a cellular-cluster scheme. 
Due to the intriguing sodium order with doping, a realistic, numerically feasible 
modeling for a large part of the phase diagram asks for a simple, but still
non-trivial approximate treatment of the structural details. We showed that 
non-local correlation effects (at least in a short-range regime) are indispensable 
to account for the phase behavior in the magnetically active region $x$$>$0.5. 
Site-dependent self-energies, triggered by the sodium ordering, are at minimum 
needed to allow for reasonable insight into the physics of Na$_x$CoO$_2$.

Emphasis was taken to study the interplay between band-structure effects and
local cluster correlations in the investigation of the highlighting magnetic 
properties. It became clear that though the van-Hove singularity at the upper
band edge (due to long-range hopping terms) is a vital ingredient, the apparent
many-body correlations on the frustrated lattice are an additional relevant
source for the enhanced susceptibilities. In this respect, the slave-boson
amplitudes for the triangular cluster multiplets proved to be valuable observables
to substantiate the latter statement. For instance, FM tendencies for Na$_x$CoO$_2$ ($t$$<$0) close
to $x$=2/3 in contrast to the strong FM susceptibility for $t$$>$0 close to $x$=1/3
may directly be connected to the local triangular cluster correlations. The
relevance of five-particle doublet cluster states in the neighborhood of 
$x$=2/3 should motivate a different viewpoint on the basic entities of the 
cobaltates. Namely, approaching the magnetic properties in this doping regime 
from a larger (not necessarily triangular) cluster perspective appears superior
to the incommensurate single-site starting point.

Aside from the simplified Stoner approach there are various more specialized 
considerations concerning ferromagnetism in Hubbard(-like) models, e.g., ideas
based on Nagaoka's theorem~\cite{nag66}, Kanamori's criterion~\cite{kan63} in the
low-density regime or ring-exchange mechanisms relevant for triangular 
plaquettes~\cite{tas98,pen96}. A very direct connection to such model ideas is
complicated because of the difficult structural facts, however it may well be that
certain aspects therefrom also apply to sodium cobaltates~\cite{mer09}. 

Our calculations revealed that the doping region 0.6$<$$x$$<$3/4 is highly 
susceptible for magnetic phase separation, pointing towards a first-order-transition
scenario for the onset of the A-type AFM order. These theoretical findings are in
accordance with many experimental studies in that 
region~\cite{muk04,sak06,bal08,zor10}, which, e.g., revealed hints for 
disorder-induced non-Fermi-liquid behavior with possibly underlying quantum 
Griffiths phases~\cite{zor10}. The region close to the band-insulating $x$=1 limit 
appears interesting for several reasons, since it is expected that a minimal hole 
doping of that state~\cite{gao07} will result in nontrivial charge susceptibilities 
compared to standard weakly correlated systems. However the simple structural model 
prohibits at this level detailed deeper investigations.

The final study on the ambitious modeling of dopings close to $x$=2/3 based
on the more elaborate effective kagom{\'e} lattice within a charge-ordered
background seem rather fruitful concerning the understanding of the interplay
between the charge and spin degrees of freedom. It showed that the charge ordering may
be efficient in singling out distinct phases that can give rise to enhanced
FM tendencies at specific dopings. This viewpoint delivers a rather natural 
possible explanation not only for the onset of FM order {\sl above} $x$=2/3, but
provides a clear-cut interpretation for the strong correlations in that doping
regime. The blocking of some of the charge not solely reduces the effective 
hopping. It also gives rise to sizeable inter-site Coulomb interactions that
strongly renormalize the remaining half-filled effective state emerging from
the neigborhood to an fluctuating CDW instability on the still accessible lattice
sites. This general idea carries over to the specific kagom\'e pattern and may
be generalized to the even higher doping regions by thinking of {\sl larger} clusters
with {\sl more} blocked sites.
Future work shall also be focused on the understanding of the spectral
properties of such phases, probably also relevant in the misfit cobaltates,
in order to touch base with already available angle-resolved photoemission (ARPES)
data~\cite{gec07,yan07,qia06,bro07,nic10}. In this respect, magnetic and ARPES 
measurements already suggested a fluctuating character of a hidden 
collective CDW-like instability~\cite{mot03,qia06}, possibly driven by long-range 
Coulomb interactions.

In summary, because of its compelling physical richness the sodium cobaltate 
system serves as a demanding testing ground for the nowadays available realistic
many-body tools. Non-local correlations effects naturally show up in Na$_x$CoO$_2$ 
and this shall stimulate further work on schemes to handle those as well as on 
many-body Hamiltonians beyond the standard Hubbard model. 

\begin{acknowledgments}
The authors are indebted to Henri Alloul, Veronique Brouet, 
Daniel Grieger and Oleg Peil for helpful disucssions. Financial support from the 
DFG-SPP 1386 is acknowledged. This research was supported in part by the National 
Science Foundation under Grant No. NSF PHY05-51164. Computations were performed at 
the local computing center of the University of Hamburg as well as the 
North-German Supercomputing Alliance (HLRN) under the grant hhp00026.
\end{acknowledgments}

\bibliographystyle{apsrev}
\bibliography{bibextra}

\begin{thebibliography}{83}
\expandafter\ifx\csname natexlab\endcsname\relax\def\natexlab#1{#1}\fi
\expandafter\ifx\csname bibnamefont\endcsname\relax
  \def\bibnamefont#1{#1}\fi
\expandafter\ifx\csname bibfnamefont\endcsname\relax
  \def\bibfnamefont#1{#1}\fi
\expandafter\ifx\csname citenamefont\endcsname\relax
  \def\citenamefont#1{#1}\fi
\expandafter\ifx\csname url\endcsname\relax
  \def\url#1{\texttt{#1}}\fi
\expandafter\ifx\csname urlprefix\endcsname\relax\def\urlprefix{URL }\fi
\providecommand{\bibinfo}[2]{#2}
\providecommand{\eprint}[2][]{\url{#2}}

\bibitem[{\citenamefont{Diep}(2005)}]{die05}
\bibinfo{author}{\bibfnamefont{H.~T.} \bibnamefont{Diep}},
  \emph{\bibinfo{title}{Frustrated Spin Systems}} (\bibinfo{publisher}{World
  Scientific Publishing Company}, \bibinfo{year}{2005}).

\bibitem[{\citenamefont{Lacroix}(2010)}]{lac10}
\bibinfo{author}{\bibfnamefont{C.}~\bibnamefont{Lacroix}}, \bibinfo{journal}{J.
  Phys. Soc. Jpn.} \textbf{\bibinfo{volume}{79}}, \bibinfo{pages}{011008}
  (\bibinfo{year}{2010}).

\bibitem[{\citenamefont{Kamihara et~al.}(2008)\citenamefont{Kamihara, Watanabe,
  Hirano, and Hosono}}]{kam08}
\bibinfo{author}{\bibfnamefont{Y.}~\bibnamefont{Kamihara}},
  \bibinfo{author}{\bibfnamefont{T.}~\bibnamefont{Watanabe}},
  \bibinfo{author}{\bibfnamefont{M.}~\bibnamefont{Hirano}}, \bibnamefont{and}
  \bibinfo{author}{\bibfnamefont{H.}~\bibnamefont{Hosono}},
  \bibinfo{journal}{J. Am. Chem. Soc.} \textbf{\bibinfo{volume}{130}},
  \bibinfo{pages}{3296} (\bibinfo{year}{2008}).

\bibitem[{\citenamefont{Si and Abrahams}(2008)}]{si08}
\bibinfo{author}{\bibfnamefont{Q.}~\bibnamefont{Si}} \bibnamefont{and}
  \bibinfo{author}{\bibfnamefont{E.}~\bibnamefont{Abrahams}},
  \bibinfo{journal}{Phys. Rev. Lett.} \textbf{\bibinfo{volume}{101}},
  \bibinfo{pages}{076401} (\bibinfo{year}{2008}).

\bibitem[{\citenamefont{Kr\"uger et~al.}(2009)\citenamefont{Kr\"uger, Kumar,
  Zaanen, and van~den Brink}}]{kru09}
\bibinfo{author}{\bibfnamefont{F.}~\bibnamefont{Kr\"uger}},
  \bibinfo{author}{\bibfnamefont{S.}~\bibnamefont{Kumar}},
  \bibinfo{author}{\bibfnamefont{J.}~\bibnamefont{Zaanen}}, \bibnamefont{and}
  \bibinfo{author}{\bibfnamefont{J.}~\bibnamefont{van~den Brink}},
  \bibinfo{journal}{Phys. Rev. B} \textbf{\bibinfo{volume}{79}},
  \bibinfo{pages}{054504} (\bibinfo{year}{2009}).

\bibitem[{\citenamefont{Lechermann et~al.}(2005)\citenamefont{Lechermann,
  Biermann, and Georges}}]{lecproc}
\bibinfo{author}{\bibfnamefont{F.}~\bibnamefont{Lechermann}},
  \bibinfo{author}{\bibfnamefont{S.}~\bibnamefont{Biermann}}, \bibnamefont{and}
  \bibinfo{author}{\bibfnamefont{A.}~\bibnamefont{Georges}},
  \bibinfo{journal}{Progress of Theoretical Physics Supplement}
  \textbf{\bibinfo{volume}{160}}, \bibinfo{pages}{233} (\bibinfo{year}{2005}).

\bibitem[{\citenamefont{Pillay et~al.}(2008)\citenamefont{Pillay, Johannes,
  Mazin, and Andersen}}]{pil08}
\bibinfo{author}{\bibfnamefont{D.}~\bibnamefont{Pillay}},
  \bibinfo{author}{\bibfnamefont{M.~D.} \bibnamefont{Johannes}},
  \bibinfo{author}{\bibfnamefont{I.~I.} \bibnamefont{Mazin}}, \bibnamefont{and}
  \bibinfo{author}{\bibfnamefont{O.~K.} \bibnamefont{Andersen}},
  \bibinfo{journal}{Phys. Rev. B} \textbf{\bibinfo{volume}{78}},
  \bibinfo{pages}{012501} (\bibinfo{year}{2008}).

\bibitem[{\citenamefont{Marianetti et~al.}(2007)\citenamefont{Marianetti,
  Haule, and Parcollet}}]{mar07}
\bibinfo{author}{\bibfnamefont{C.~A.} \bibnamefont{Marianetti}},
  \bibinfo{author}{\bibfnamefont{K.}~\bibnamefont{Haule}}, \bibnamefont{and}
  \bibinfo{author}{\bibfnamefont{O.}~\bibnamefont{Parcollet}},
  \bibinfo{journal}{Phys. Rev. Lett.} \textbf{\bibinfo{volume}{99}},
  \bibinfo{pages}{246404} (\bibinfo{year}{2007}).

\bibitem[{\citenamefont{Wang et~al.}(2008)\citenamefont{Wang, Dai, and
  Fang}}]{wan08}
\bibinfo{author}{\bibfnamefont{G.-T.} \bibnamefont{Wang}},
  \bibinfo{author}{\bibfnamefont{X.}~\bibnamefont{Dai}}, \bibnamefont{and}
  \bibinfo{author}{\bibfnamefont{Z.}~\bibnamefont{Fang}},
  \bibinfo{journal}{Phys. Rev. Lett.} \textbf{\bibinfo{volume}{101}},
  \bibinfo{pages}{066403} (\bibinfo{year}{2008}).

\bibitem[{\citenamefont{Liebsch and Ishida}(2008)}]{lie08}
\bibinfo{author}{\bibfnamefont{A.}~\bibnamefont{Liebsch}} \bibnamefont{and}
  \bibinfo{author}{\bibfnamefont{H.}~\bibnamefont{Ishida}},
  \bibinfo{journal}{Eur. Phys. J. B} \textbf{\bibinfo{volume}{61}},
  \bibinfo{pages}{405} (\bibinfo{year}{2008}).

\bibitem[{\citenamefont{Bourgeois et~al.}(2009)\citenamefont{Bourgeois, Aligia,
  and Rozenberg}}]{bou09}
\bibinfo{author}{\bibfnamefont{A.}~\bibnamefont{Bourgeois}},
  \bibinfo{author}{\bibfnamefont{A.~A.} \bibnamefont{Aligia}},
  \bibnamefont{and} \bibinfo{author}{\bibfnamefont{M.~J.}
  \bibnamefont{Rozenberg}}, \bibinfo{journal}{Phys. Rev. Lett.}
  \textbf{\bibinfo{volume}{102}}, \bibinfo{pages}{066402}
  (\bibinfo{year}{2009}).

\bibitem[{\citenamefont{Takada et~al.}(2003)\citenamefont{Takada, Sakurai,
  Takayama-Muromachi, Izumi, Dilanian, and Sasaki}}]{tak03}
\bibinfo{author}{\bibfnamefont{K.}~\bibnamefont{Takada}},
  \bibinfo{author}{\bibfnamefont{H.}~\bibnamefont{Sakurai}},
  \bibinfo{author}{\bibfnamefont{E.}~\bibnamefont{Takayama-Muromachi}},
  \bibinfo{author}{\bibfnamefont{F.}~\bibnamefont{Izumi}},
  \bibinfo{author}{\bibfnamefont{R.~A.} \bibnamefont{Dilanian}},
  \bibnamefont{and} \bibinfo{author}{\bibfnamefont{T.}~\bibnamefont{Sasaki}},
  \bibinfo{journal}{Nature} \textbf{\bibinfo{volume}{422}}, \bibinfo{pages}{53}
  (\bibinfo{year}{2003}).

\bibitem[{\citenamefont{Foo et~al.}(2004)\citenamefont{Foo, Wang, Watauchi,
  Zandbergen, He, Cava, and Ong}}]{foo04}
\bibinfo{author}{\bibfnamefont{M.~L.} \bibnamefont{Foo}},
  \bibinfo{author}{\bibfnamefont{Y.}~\bibnamefont{Wang}},
  \bibinfo{author}{\bibfnamefont{S.}~\bibnamefont{Watauchi}},
  \bibinfo{author}{\bibfnamefont{H.~W.} \bibnamefont{Zandbergen}},
  \bibinfo{author}{\bibfnamefont{T.}~\bibnamefont{He}},
  \bibinfo{author}{\bibfnamefont{R.~J.} \bibnamefont{Cava}}, \bibnamefont{and}
  \bibinfo{author}{\bibfnamefont{N.~P.} \bibnamefont{Ong}},
  \bibinfo{journal}{Phys. Rev. Lett.} \textbf{\bibinfo{volume}{92}},
  \bibinfo{pages}{247001} (\bibinfo{year}{2004}).

\bibitem[{\citenamefont{Mukhamedshin et~al.}(2005)\citenamefont{Mukhamedshin,
  Alloul, Collin, and Blanchard}}]{muk05}
\bibinfo{author}{\bibfnamefont{I.~R.} \bibnamefont{Mukhamedshin}},
  \bibinfo{author}{\bibfnamefont{H.}~\bibnamefont{Alloul}},
  \bibinfo{author}{\bibfnamefont{G.}~\bibnamefont{Collin}}, \bibnamefont{and}
  \bibinfo{author}{\bibfnamefont{N.}~\bibnamefont{Blanchard}},
  \bibinfo{journal}{Phys. Rev. Lett.} \textbf{\bibinfo{volume}{94}},
  \bibinfo{pages}{247602} (\bibinfo{year}{2005}).

\bibitem[{\citenamefont{Lang et~al.}(2008)\citenamefont{Lang, Bobroff, Alloul,
  Collin, and Blanchard}}]{lan08}
\bibinfo{author}{\bibfnamefont{G.}~\bibnamefont{Lang}},
  \bibinfo{author}{\bibfnamefont{J.}~\bibnamefont{Bobroff}},
  \bibinfo{author}{\bibfnamefont{H.}~\bibnamefont{Alloul}},
  \bibinfo{author}{\bibfnamefont{G.}~\bibnamefont{Collin}}, \bibnamefont{and}
  \bibinfo{author}{\bibfnamefont{N.}~\bibnamefont{Blanchard}},
  \bibinfo{journal}{Phys. Rev. B} \textbf{\bibinfo{volume}{78}},
  \bibinfo{pages}{155116} (\bibinfo{year}{2008}).

\bibitem[{\citenamefont{Wang et~al.}(2003)\citenamefont{Wang, Rogado, Cava, and
  Ong}}]{wan03}
\bibinfo{author}{\bibfnamefont{Y.}~\bibnamefont{Wang}},
  \bibinfo{author}{\bibfnamefont{N.~S.} \bibnamefont{Rogado}},
  \bibinfo{author}{\bibfnamefont{R.~J.} \bibnamefont{Cava}}, \bibnamefont{and}
  \bibinfo{author}{\bibfnamefont{N.~P.} \bibnamefont{Ong}},
  \bibinfo{journal}{Nature} \textbf{\bibinfo{volume}{423}},
  \bibinfo{pages}{425} (\bibinfo{year}{2003}).

\bibitem[{\citenamefont{Fujimoto et~al.}(2004)\citenamefont{Fujimoto, Zheng,
  Kitaoka, Meng, Cmaidalka, and Chu}}]{fuj04}
\bibinfo{author}{\bibfnamefont{T.}~\bibnamefont{Fujimoto}},
  \bibinfo{author}{\bibfnamefont{G.-Q.} \bibnamefont{Zheng}},
  \bibinfo{author}{\bibfnamefont{Y.}~\bibnamefont{Kitaoka}},
  \bibinfo{author}{\bibfnamefont{R.~L.} \bibnamefont{Meng}},
  \bibinfo{author}{\bibfnamefont{J.}~\bibnamefont{Cmaidalka}},
  \bibnamefont{and} \bibinfo{author}{\bibfnamefont{C.~W.} \bibnamefont{Chu}},
  \bibinfo{journal}{Phys. Rev. Lett.} \textbf{\bibinfo{volume}{92}},
  \bibinfo{pages}{047004} (\bibinfo{year}{2004}).

\bibitem[{\citenamefont{Yokoi et~al.}(2005)\citenamefont{Yokoi, Moyoshi,
  Kobayashi, Soda, Yasui, Sato, and Kakurai}}]{yok05}
\bibinfo{author}{\bibfnamefont{M.}~\bibnamefont{Yokoi}},
  \bibinfo{author}{\bibfnamefont{T.}~\bibnamefont{Moyoshi}},
  \bibinfo{author}{\bibfnamefont{Y.}~\bibnamefont{Kobayashi}},
  \bibinfo{author}{\bibfnamefont{M.}~\bibnamefont{Soda}},
  \bibinfo{author}{\bibfnamefont{Y.}~\bibnamefont{Yasui}},
  \bibinfo{author}{\bibfnamefont{M.}~\bibnamefont{Sato}}, \bibnamefont{and}
  \bibinfo{author}{\bibfnamefont{K.}~\bibnamefont{Kakurai}},
  \bibinfo{journal}{J. Phys. Soc. Jpn.} \textbf{\bibinfo{volume}{74}},
  \bibinfo{pages}{3046} (\bibinfo{year}{2005}).

\bibitem[{\citenamefont{Kawasaki et~al.}(2009)\citenamefont{Kawasaki,
  Motohashi, Shimada, Ono, Kanno, Karppinen, Yamauchi, , and Zheng}}]{kaw09}
\bibinfo{author}{\bibfnamefont{S.}~\bibnamefont{Kawasaki}},
  \bibinfo{author}{\bibfnamefont{T.}~\bibnamefont{Motohashi}},
  \bibinfo{author}{\bibfnamefont{K.}~\bibnamefont{Shimada}},
  \bibinfo{author}{\bibfnamefont{T.}~\bibnamefont{Ono}},
  \bibinfo{author}{\bibfnamefont{R.}~\bibnamefont{Kanno}},
  \bibinfo{author}{\bibfnamefont{M.}~\bibnamefont{Karppinen}},
  \bibinfo{author}{\bibfnamefont{H.}~\bibnamefont{Yamauchi}}, ,
  \bibnamefont{and} \bibinfo{author}{\bibfnamefont{G.-Q.} \bibnamefont{Zheng}},
  \bibinfo{journal}{Phys. Rev. B} \textbf{\bibinfo{volume}{79}},
  \bibinfo{pages}{220514} (\bibinfo{year}{2009}).

\bibitem[{\citenamefont{de~Vaulx et~al.}(2007)\citenamefont{de~Vaulx, Julien,
  Berthier, H\'{e}bert, Pralong, and Maignan}}]{vau07}
\bibinfo{author}{\bibfnamefont{C.}~\bibnamefont{de~Vaulx}},
  \bibinfo{author}{\bibfnamefont{M.-H.} \bibnamefont{Julien}},
  \bibinfo{author}{\bibfnamefont{C.}~\bibnamefont{Berthier}},
  \bibinfo{author}{\bibfnamefont{S.}~\bibnamefont{H\'{e}bert}},
  \bibinfo{author}{\bibfnamefont{V.}~\bibnamefont{Pralong}}, \bibnamefont{and}
  \bibinfo{author}{\bibfnamefont{A.}~\bibnamefont{Maignan}},
  \bibinfo{journal}{Phys. Rev. Lett.} \textbf{\bibinfo{volume}{98}},
  \bibinfo{pages}{246402} (\bibinfo{year}{2007}).

\bibitem[{\citenamefont{Mendels et~al.}(2005)\citenamefont{Mendels, Bono,
  Bobroff, Collin, Colson, Blanchard, Alloul, Mukhamedshin, Bert, Amato
  et~al.}}]{men05}
\bibinfo{author}{\bibfnamefont{P.}~\bibnamefont{Mendels}},
  \bibinfo{author}{\bibfnamefont{D.}~\bibnamefont{Bono}},
  \bibinfo{author}{\bibfnamefont{J.}~\bibnamefont{Bobroff}},
  \bibinfo{author}{\bibfnamefont{G.}~\bibnamefont{Collin}},
  \bibinfo{author}{\bibfnamefont{D.}~\bibnamefont{Colson}},
  \bibinfo{author}{\bibfnamefont{N.}~\bibnamefont{Blanchard}},
  \bibinfo{author}{\bibfnamefont{H.}~\bibnamefont{Alloul}},
  \bibinfo{author}{\bibfnamefont{I.}~\bibnamefont{Mukhamedshin}},
  \bibinfo{author}{\bibfnamefont{F.}~\bibnamefont{Bert}},
  \bibinfo{author}{\bibfnamefont{A.}~\bibnamefont{Amato}},
  \bibnamefont{et~al.}, \bibinfo{journal}{Phys. Rev. Lett.}
  \textbf{\bibinfo{volume}{94}}, \bibinfo{pages}{136403}
  (\bibinfo{year}{2005}).

\bibitem[{\citenamefont{Sugiyama et~al.}(2003)\citenamefont{Sugiyama, Itahara,
  Brewer, Ansaldo, Motohashi, Karppinen, and Yamauchi}}]{sug03}
\bibinfo{author}{\bibfnamefont{J.}~\bibnamefont{Sugiyama}},
  \bibinfo{author}{\bibfnamefont{H.}~\bibnamefont{Itahara}},
  \bibinfo{author}{\bibfnamefont{J.~H.} \bibnamefont{Brewer}},
  \bibinfo{author}{\bibfnamefont{E.~J.} \bibnamefont{Ansaldo}},
  \bibinfo{author}{\bibfnamefont{T.}~\bibnamefont{Motohashi}},
  \bibinfo{author}{\bibfnamefont{M.}~\bibnamefont{Karppinen}},
  \bibnamefont{and} \bibinfo{author}{\bibfnamefont{H.}~\bibnamefont{Yamauchi}},
  \bibinfo{journal}{Phys. Rev. B} \textbf{\bibinfo{volume}{67}},
  \bibinfo{pages}{214420} (\bibinfo{year}{2003}).

\bibitem[{\citenamefont{Motohashi et~al.}(2003)\citenamefont{Motohashi, Ueda,
  Naujalis, Tojo, Terasaki, Atake, Karppinen, and Yamauchi}}]{mot03}
\bibinfo{author}{\bibfnamefont{T.}~\bibnamefont{Motohashi}},
  \bibinfo{author}{\bibfnamefont{R.}~\bibnamefont{Ueda}},
  \bibinfo{author}{\bibfnamefont{E.}~\bibnamefont{Naujalis}},
  \bibinfo{author}{\bibfnamefont{T.}~\bibnamefont{Tojo}},
  \bibinfo{author}{\bibfnamefont{I.}~\bibnamefont{Terasaki}},
  \bibinfo{author}{\bibfnamefont{T.}~\bibnamefont{Atake}},
  \bibinfo{author}{\bibfnamefont{M.}~\bibnamefont{Karppinen}},
  \bibnamefont{and} \bibinfo{author}{\bibfnamefont{H.}~\bibnamefont{Yamauchi}},
  \bibinfo{journal}{Phys. Rev. B} \textbf{\bibinfo{volume}{67}},
  \bibinfo{pages}{064406} (\bibinfo{year}{2003}).

\bibitem[{\citenamefont{Boothroyd et~al.}(2004)\citenamefont{Boothroyd, Coldea,
  Tennant, Prabhakaran, Helme, and Frost}}]{boo04}
\bibinfo{author}{\bibfnamefont{A.~T.} \bibnamefont{Boothroyd}},
  \bibinfo{author}{\bibfnamefont{R.}~\bibnamefont{Coldea}},
  \bibinfo{author}{\bibfnamefont{D.~A.} \bibnamefont{Tennant}},
  \bibinfo{author}{\bibfnamefont{D.}~\bibnamefont{Prabhakaran}},
  \bibinfo{author}{\bibfnamefont{L.~M.} \bibnamefont{Helme}}, \bibnamefont{and}
  \bibinfo{author}{\bibfnamefont{C.~D.} \bibnamefont{Frost}},
  \bibinfo{journal}{Phys. Rev. Lett.} \textbf{\bibinfo{volume}{92}},
  \bibinfo{pages}{197201} (\bibinfo{year}{2004}).

\bibitem[{\citenamefont{Ihara et~al.}(2004)\citenamefont{Ihara, Ishida,
  Michioka, Kato, Yoshimura1, Sakurai, and Takayama-Muromachi}}]{iha04}
\bibinfo{author}{\bibfnamefont{Y.}~\bibnamefont{Ihara}},
  \bibinfo{author}{\bibfnamefont{K.}~\bibnamefont{Ishida}},
  \bibinfo{author}{\bibfnamefont{C.}~\bibnamefont{Michioka}},
  \bibinfo{author}{\bibfnamefont{M.}~\bibnamefont{Kato}},
  \bibinfo{author}{\bibfnamefont{K.}~\bibnamefont{Yoshimura1}},
  \bibinfo{author}{\bibfnamefont{H.}~\bibnamefont{Sakurai}}, \bibnamefont{and}
  \bibinfo{author}{\bibfnamefont{E.}~\bibnamefont{Takayama-Muromachi}},
  \bibinfo{journal}{J. Phys. Soc. Jpn.} \textbf{\bibinfo{volume}{73}},
  \bibinfo{pages}{2963} (\bibinfo{year}{2004}).

\bibitem[{\citenamefont{Sakurai et~al.}(2004)\citenamefont{Sakurai, Tsujii, and
  Takayama-Muromachi}}]{sak04}
\bibinfo{author}{\bibfnamefont{H.}~\bibnamefont{Sakurai}},
  \bibinfo{author}{\bibfnamefont{N.}~\bibnamefont{Tsujii}}, \bibnamefont{and}
  \bibinfo{author}{\bibfnamefont{E.}~\bibnamefont{Takayama-Muromachi}},
  \bibinfo{journal}{J. Phys. Soc. Jpn.} \textbf{\bibinfo{volume}{73}},
  \bibinfo{pages}{2393} (\bibinfo{year}{2004}).

\bibitem[{\citenamefont{Bayrakci et~al.}(2005)\citenamefont{Bayrakci, Mirebeau,
  Bourges, Sidis, Enderle, Mesot, Chen, Lin, and Keimer}}]{bay05}
\bibinfo{author}{\bibfnamefont{S.~P.} \bibnamefont{Bayrakci}},
  \bibinfo{author}{\bibfnamefont{I.}~\bibnamefont{Mirebeau}},
  \bibinfo{author}{\bibfnamefont{P.}~\bibnamefont{Bourges}},
  \bibinfo{author}{\bibfnamefont{Y.}~\bibnamefont{Sidis}},
  \bibinfo{author}{\bibfnamefont{M.}~\bibnamefont{Enderle}},
  \bibinfo{author}{\bibfnamefont{J.}~\bibnamefont{Mesot}},
  \bibinfo{author}{\bibfnamefont{D.~P.} \bibnamefont{Chen}},
  \bibinfo{author}{\bibfnamefont{C.~T.} \bibnamefont{Lin}}, \bibnamefont{and}
  \bibinfo{author}{\bibfnamefont{B.}~\bibnamefont{Keimer}},
  \bibinfo{journal}{Phys. Rev. Lett.} \textbf{\bibinfo{volume}{94}},
  \bibinfo{pages}{157205} (\bibinfo{year}{2005}).

\bibitem[{\citenamefont{Helme et~al.}(2006)\citenamefont{Helme, Boothroyd,
  Coldea, Prabhakaran, Stunault, McIntyre, and Kernavanois}}]{hel06}
\bibinfo{author}{\bibfnamefont{L.~M.} \bibnamefont{Helme}},
  \bibinfo{author}{\bibfnamefont{A.~T.} \bibnamefont{Boothroyd}},
  \bibinfo{author}{\bibfnamefont{R.}~\bibnamefont{Coldea}},
  \bibinfo{author}{\bibfnamefont{D.}~\bibnamefont{Prabhakaran}},
  \bibinfo{author}{\bibfnamefont{A.}~\bibnamefont{Stunault}},
  \bibinfo{author}{\bibfnamefont{G.~J.} \bibnamefont{McIntyre}},
  \bibnamefont{and}
  \bibinfo{author}{\bibfnamefont{N.}~\bibnamefont{Kernavanois}},
  \bibinfo{journal}{Phys. Rev. B} \textbf{\bibinfo{volume}{73}},
  \bibinfo{pages}{054405} (\bibinfo{year}{2006}).

\bibitem[{\citenamefont{Shu et~al.}(2007)\citenamefont{Shu, Prodi, Chu, Lee,
  Sheu, and Chou}}]{shu07}
\bibinfo{author}{\bibfnamefont{G.~J.} \bibnamefont{Shu}},
  \bibinfo{author}{\bibfnamefont{A.}~\bibnamefont{Prodi}},
  \bibinfo{author}{\bibfnamefont{S.~Y.} \bibnamefont{Chu}},
  \bibinfo{author}{\bibfnamefont{Y.~S.} \bibnamefont{Lee}},
  \bibinfo{author}{\bibfnamefont{S.}~\bibnamefont{Sheu}}, \bibnamefont{and}
  \bibinfo{author}{\bibfnamefont{F.~C.} \bibnamefont{Chou}},
  \bibinfo{journal}{Phys. Rev. B} \textbf{\bibinfo{volume}{76}},
  \bibinfo{pages}{184115} (\bibinfo{year}{2007}).

\bibitem[{\citenamefont{Schulze et~al.}(2008)\citenamefont{Schulze,
  Br{\"u}hwiler, H{\"a}fliger, Kazakov, Niedermayer, Mattenberger, Karpinski,
  and Batlogg}}]{schu08}
\bibinfo{author}{\bibfnamefont{T.~F.} \bibnamefont{Schulze}},
  \bibinfo{author}{\bibfnamefont{M.}~\bibnamefont{Br{\"u}hwiler}},
  \bibinfo{author}{\bibfnamefont{P.~S.} \bibnamefont{H{\"a}fliger}},
  \bibinfo{author}{\bibfnamefont{S.~M.} \bibnamefont{Kazakov}},
  \bibinfo{author}{\bibfnamefont{C.}~\bibnamefont{Niedermayer}},
  \bibinfo{author}{\bibfnamefont{K.}~\bibnamefont{Mattenberger}},
  \bibinfo{author}{\bibfnamefont{J.}~\bibnamefont{Karpinski}},
  \bibnamefont{and} \bibinfo{author}{\bibfnamefont{B.}~\bibnamefont{Batlogg}},
  \bibinfo{journal}{Phys. Rev. B} \textbf{\bibinfo{volume}{78}},
  \bibinfo{pages}{205101} (\bibinfo{year}{2008}).

\bibitem[{\citenamefont{Singh}(2000)}]{sin00}
\bibinfo{author}{\bibfnamefont{D.}~\bibnamefont{Singh}},
  \bibinfo{journal}{Phys. Rev. B} \textbf{\bibinfo{volume}{61}},
  \bibinfo{pages}{13397} (\bibinfo{year}{2000}).

\bibitem[{\citenamefont{Singh}(2003)}]{sin03}
\bibinfo{author}{\bibfnamefont{D.}~\bibnamefont{Singh}},
  \bibinfo{journal}{Phys. Rev. B} \textbf{\bibinfo{volume}{68}},
  \bibinfo{pages}{020503} (\bibinfo{year}{2003}).

\bibitem[{\citenamefont{Johannes et~al.}(2005)\citenamefont{Johannes, Mazin,
  and Singh}}]{joh05}
\bibinfo{author}{\bibfnamefont{M.~D.} \bibnamefont{Johannes}},
  \bibinfo{author}{\bibfnamefont{I.~I.} \bibnamefont{Mazin}}, \bibnamefont{and}
  \bibinfo{author}{\bibfnamefont{D.~J.} \bibnamefont{Singh}},
  \bibinfo{journal}{Phys. Rev. B} \textbf{\bibinfo{volume}{71}},
  \bibinfo{pages}{214410} (\bibinfo{year}{2005}).

\bibitem[{\citenamefont{Wang et~al.}(2004)\citenamefont{Wang, Zheng, Wu, Ma,
  Xiang, Jin, and Mandrus}}]{wan04}
\bibinfo{author}{\bibfnamefont{N.~L.} \bibnamefont{Wang}},
  \bibinfo{author}{\bibfnamefont{P.}~\bibnamefont{Zheng}},
  \bibinfo{author}{\bibfnamefont{D.}~\bibnamefont{Wu}},
  \bibinfo{author}{\bibfnamefont{Y.~C.} \bibnamefont{Ma}},
  \bibinfo{author}{\bibfnamefont{T.}~\bibnamefont{Xiang}},
  \bibinfo{author}{\bibfnamefont{R.~Y.} \bibnamefont{Jin}}, \bibnamefont{and}
  \bibinfo{author}{\bibfnamefont{D.}~\bibnamefont{Mandrus}},
  \bibinfo{journal}{Phys. Rev. Lett.} \textbf{\bibinfo{volume}{93}},
  \bibinfo{pages}{237007} (\bibinfo{year}{2004}).

\bibitem[{\citenamefont{Valla et~al.}(2002)\citenamefont{Valla, Johnson, Yusof,
  Wells, Li, Loureiro, Cava, Mikami, Y.~Mori, and Sasaki}}]{val02}
\bibinfo{author}{\bibfnamefont{T.}~\bibnamefont{Valla}},
  \bibinfo{author}{\bibfnamefont{P.~D.} \bibnamefont{Johnson}},
  \bibinfo{author}{\bibfnamefont{Z.}~\bibnamefont{Yusof}},
  \bibinfo{author}{\bibfnamefont{B.}~\bibnamefont{Wells}},
  \bibinfo{author}{\bibfnamefont{Q.}~\bibnamefont{Li}},
  \bibinfo{author}{\bibfnamefont{S.~M.} \bibnamefont{Loureiro}},
  \bibinfo{author}{\bibfnamefont{R.~J.} \bibnamefont{Cava}},
  \bibinfo{author}{\bibfnamefont{M.}~\bibnamefont{Mikami}},
  \bibinfo{author}{\bibfnamefont{M.~Y.} \bibnamefont{Y.~Mori}},
  \bibnamefont{and} \bibinfo{author}{\bibfnamefont{T.}~\bibnamefont{Sasaki}},
  \bibinfo{journal}{Nature} \textbf{\bibinfo{volume}{417}},
  \bibinfo{pages}{627} (\bibinfo{year}{2002}).

\bibitem[{\citenamefont{Hasan et~al.}(2004)\citenamefont{Hasan, Chuang, Qian,
  Li, Kong, Kuprin, Fedorov, Kimmerling, Rotenberg, Rossnagel et~al.}}]{has04}
\bibinfo{author}{\bibfnamefont{M.~Z.} \bibnamefont{Hasan}},
  \bibinfo{author}{\bibfnamefont{Y.-D.} \bibnamefont{Chuang}},
  \bibinfo{author}{\bibfnamefont{D.}~\bibnamefont{Qian}},
  \bibinfo{author}{\bibfnamefont{Y.~W.} \bibnamefont{Li}},
  \bibinfo{author}{\bibfnamefont{Y.}~\bibnamefont{Kong}},
  \bibinfo{author}{\bibfnamefont{A.}~\bibnamefont{Kuprin}},
  \bibinfo{author}{\bibfnamefont{A.~V.} \bibnamefont{Fedorov}},
  \bibinfo{author}{\bibfnamefont{R.}~\bibnamefont{Kimmerling}},
  \bibinfo{author}{\bibfnamefont{E.}~\bibnamefont{Rotenberg}},
  \bibinfo{author}{\bibfnamefont{K.}~\bibnamefont{Rossnagel}},
  \bibnamefont{et~al.}, \bibinfo{journal}{Phys. Rev. Lett.}
  \textbf{\bibinfo{volume}{92}}, \bibinfo{pages}{246402}
  (\bibinfo{year}{2004}).

\bibitem[{\citenamefont{Yang et~al.}(2007)\citenamefont{Yang, Wang, and
  Ding}}]{yan07}
\bibinfo{author}{\bibfnamefont{H.-B.} \bibnamefont{Yang}},
  \bibinfo{author}{\bibfnamefont{Z.}~\bibnamefont{Wang}}, \bibnamefont{and}
  \bibinfo{author}{\bibfnamefont{H.}~\bibnamefont{Ding}}, \bibinfo{journal}{J.
  Phys.: Condens. Matter} \textbf{\bibinfo{volume}{19}},
  \bibinfo{pages}{355004} (\bibinfo{year}{2007}).

\bibitem[{\citenamefont{Geck et~al.}(2007)\citenamefont{Geck, Borisenko,
  Berger, Eschrig, Fink, Knupfer, Koepernik, A.Koitzsch, Kordyuk, Zabolotnyy
  et~al.}}]{gec07}
\bibinfo{author}{\bibfnamefont{J.}~\bibnamefont{Geck}},
  \bibinfo{author}{\bibfnamefont{S.~V.} \bibnamefont{Borisenko}},
  \bibinfo{author}{\bibfnamefont{H.}~\bibnamefont{Berger}},
  \bibinfo{author}{\bibfnamefont{H.}~\bibnamefont{Eschrig}},
  \bibinfo{author}{\bibfnamefont{J.}~\bibnamefont{Fink}},
  \bibinfo{author}{\bibfnamefont{M.}~\bibnamefont{Knupfer}},
  \bibinfo{author}{\bibfnamefont{K.}~\bibnamefont{Koepernik}},
  \bibinfo{author}{\bibnamefont{A.Koitzsch}},
  \bibinfo{author}{\bibfnamefont{A.~A.} \bibnamefont{Kordyuk}},
  \bibinfo{author}{\bibfnamefont{V.~B.} \bibnamefont{Zabolotnyy}},
  \bibnamefont{et~al.}, \bibinfo{journal}{Phys. Rev. Lett.}
  \textbf{\bibinfo{volume}{99}}, \bibinfo{pages}{046403}
  (\bibinfo{year}{2007}).

\bibitem[{\citenamefont{Merino et~al.}(2006)\citenamefont{Merino, Powell, and
  McKenzie}}]{mer06}
\bibinfo{author}{\bibfnamefont{J.}~\bibnamefont{Merino}},
  \bibinfo{author}{\bibfnamefont{B.~J.} \bibnamefont{Powell}},
  \bibnamefont{and} \bibinfo{author}{\bibfnamefont{R.~H.}
  \bibnamefont{McKenzie}}, \bibinfo{journal}{Phys. Rev. B}
  \textbf{\bibinfo{volume}{73}}, \bibinfo{pages}{235107}
  (\bibinfo{year}{2006}).

\bibitem[{\citenamefont{Rosner et~al.}(2003)\citenamefont{Rosner, Drechsler,
  Fuchs, Handstein, W\"alte, and M\"uller}}]{ros03}
\bibinfo{author}{\bibfnamefont{H.}~\bibnamefont{Rosner}},
  \bibinfo{author}{\bibfnamefont{S.-L.} \bibnamefont{Drechsler}},
  \bibinfo{author}{\bibfnamefont{G.}~\bibnamefont{Fuchs}},
  \bibinfo{author}{\bibfnamefont{A.}~\bibnamefont{Handstein}},
  \bibinfo{author}{\bibfnamefont{A.}~\bibnamefont{W\"alte}}, \bibnamefont{and}
  \bibinfo{author}{\bibfnamefont{K.-H.} \bibnamefont{M\"uller}},
  \bibinfo{journal}{Braz. J. Phys.} \textbf{\bibinfo{volume}{33}},
  \bibinfo{pages}{718} (\bibinfo{year}{2003}).

\bibitem[{\citenamefont{Johannes et~al.}(2004)\citenamefont{Johannes,
  Papaconstantopoulos, Singh, and Mehl}}]{joh04}
\bibinfo{author}{\bibfnamefont{M.~D.} \bibnamefont{Johannes}},
  \bibinfo{author}{\bibfnamefont{D.~A.} \bibnamefont{Papaconstantopoulos}},
  \bibinfo{author}{\bibfnamefont{D.~J.} \bibnamefont{Singh}}, \bibnamefont{and}
  \bibinfo{author}{\bibfnamefont{M.~J.} \bibnamefont{Mehl}},
  \bibinfo{journal}{Europhys. Lett.} \textbf{\bibinfo{volume}{68}},
  \bibinfo{pages}{433} (\bibinfo{year}{2004}).

\bibitem[{\citenamefont{Lee et~al.}(2004)\citenamefont{Lee, Kunes, and
  Pickett}}]{lee04}
\bibinfo{author}{\bibfnamefont{K.~W.} \bibnamefont{Lee}},
  \bibinfo{author}{\bibfnamefont{J.}~\bibnamefont{Kunes}}, \bibnamefont{and}
  \bibinfo{author}{\bibfnamefont{W.~E.} \bibnamefont{Pickett}},
  \bibinfo{journal}{Phys. Rev. B} \textbf{\bibinfo{volume}{70}},
  \bibinfo{pages}{045104} (\bibinfo{year}{2004}).

\bibitem[{\citenamefont{Korshunov et~al.}(2007)\citenamefont{Korshunov, Eremin,
  Shorikov, Anisimov, Renner, and Brenig}}]{kor07}
\bibinfo{author}{\bibfnamefont{M.~M.} \bibnamefont{Korshunov}},
  \bibinfo{author}{\bibfnamefont{I.}~\bibnamefont{Eremin}},
  \bibinfo{author}{\bibfnamefont{A.}~\bibnamefont{Shorikov}},
  \bibinfo{author}{\bibfnamefont{V.~I.} \bibnamefont{Anisimov}},
  \bibinfo{author}{\bibfnamefont{M.}~\bibnamefont{Renner}}, \bibnamefont{and}
  \bibinfo{author}{\bibfnamefont{W.}~\bibnamefont{Brenig}},
  \bibinfo{journal}{Phys. Rev. B} \textbf{\bibinfo{volume}{75}},
  \bibinfo{pages}{094511} (\bibinfo{year}{2007}).

\bibitem[{\citenamefont{Gao et~al.}(2007)\citenamefont{Gao, Zhou, and
  Wang}}]{gao07}
\bibinfo{author}{\bibfnamefont{M.}~\bibnamefont{Gao}},
  \bibinfo{author}{\bibfnamefont{S.}~\bibnamefont{Zhou}}, \bibnamefont{and}
  \bibinfo{author}{\bibfnamefont{Z.}~\bibnamefont{Wang}},
  \bibinfo{journal}{Phys. Rev. B} \textbf{\bibinfo{volume}{76}},
  \bibinfo{pages}{180402} (\bibinfo{year}{2007}).

\bibitem[{\citenamefont{Nordheim}(1931)}]{nor31}
\bibinfo{author}{\bibfnamefont{L.}~\bibnamefont{Nordheim}},
  \bibinfo{journal}{Ann. Phys. (Leipzig)} \textbf{\bibinfo{volume}{9}},
  \bibinfo{pages}{607} (\bibinfo{year}{1931}).

\bibitem[{\citenamefont{Marianetti and Kotliar}(2007)}]{markot07}
\bibinfo{author}{\bibfnamefont{C.~A.} \bibnamefont{Marianetti}}
  \bibnamefont{and} \bibinfo{author}{\bibfnamefont{G.}~\bibnamefont{Kotliar}},
  \bibinfo{journal}{Phys. Rev. Lett.} \textbf{\bibinfo{volume}{98}},
  \bibinfo{pages}{176405} (\bibinfo{year}{2007}).

\bibitem[{\citenamefont{Haerter et~al.}(2006)\citenamefont{Haerter, Peterson,
  and Shastry}}]{hae06}
\bibinfo{author}{\bibfnamefont{J.~O.} \bibnamefont{Haerter}},
  \bibinfo{author}{\bibfnamefont{M.~R.} \bibnamefont{Peterson}},
  \bibnamefont{and} \bibinfo{author}{\bibfnamefont{B.~S.}
  \bibnamefont{Shastry}}, \bibinfo{journal}{Phys. Rev. Lett.}
  \textbf{\bibinfo{volume}{97}}, \bibinfo{pages}{226402}
  (\bibinfo{year}{2006}).

\bibitem[{\citenamefont{Alloul et~al.}(2009)\citenamefont{Alloul,
  Mukhamedishin, Platova, and Dooglav}}]{all09}
\bibinfo{author}{\bibfnamefont{H.}~\bibnamefont{Alloul}},
  \bibinfo{author}{\bibfnamefont{I.~R.} \bibnamefont{Mukhamedishin}},
  \bibinfo{author}{\bibfnamefont{T.~A.} \bibnamefont{Platova}},
  \bibnamefont{and} \bibinfo{author}{\bibfnamefont{A.~V.}
  \bibnamefont{Dooglav}}, \bibinfo{journal}{Europhys. Lett.}
  \textbf{\bibinfo{volume}{85}}, \bibinfo{pages}{47006} (\bibinfo{year}{2009}).

\bibitem[{\citenamefont{Lechermann}(2009)}]{lec09}
\bibinfo{author}{\bibfnamefont{F.}~\bibnamefont{Lechermann}},
  \bibinfo{journal}{Phys. Rev. Lett.} \textbf{\bibinfo{volume}{102}},
  \bibinfo{pages}{046403} (\bibinfo{year}{2009}).

\bibitem[{\citenamefont{Li et~al.}(1989)\citenamefont{Li, W\"olfle, and
  Hirschfeld}}]{li89}
\bibinfo{author}{\bibfnamefont{T.}~\bibnamefont{Li}},
  \bibinfo{author}{\bibfnamefont{P.}~\bibnamefont{W\"olfle}}, \bibnamefont{and}
  \bibinfo{author}{\bibfnamefont{P.~J.} \bibnamefont{Hirschfeld}},
  \bibinfo{journal}{Phys. Rev. B} \textbf{\bibinfo{volume}{40}},
  \bibinfo{pages}{6817} (\bibinfo{year}{1989}).

\bibitem[{\citenamefont{Lechermann et~al.}(2007)\citenamefont{Lechermann,
  Georges, Kotliar, and Parcollet}}]{lec07}
\bibinfo{author}{\bibfnamefont{F.}~\bibnamefont{Lechermann}},
  \bibinfo{author}{\bibfnamefont{A.}~\bibnamefont{Georges}},
  \bibinfo{author}{\bibfnamefont{G.}~\bibnamefont{Kotliar}}, \bibnamefont{and}
  \bibinfo{author}{\bibfnamefont{O.}~\bibnamefont{Parcollet}},
  \bibinfo{journal}{Phys. Rev. B} \textbf{\bibinfo{volume}{76}},
  \bibinfo{pages}{155102} (\bibinfo{year}{2007}).

\bibitem[{\citenamefont{Galanakis et~al.}(2009)\citenamefont{Galanakis,
  Stanescu, and Phillips}}]{gal09}
\bibinfo{author}{\bibfnamefont{D.}~\bibnamefont{Galanakis}},
  \bibinfo{author}{\bibfnamefont{T.~D.} \bibnamefont{Stanescu}},
  \bibnamefont{and} \bibinfo{author}{\bibfnamefont{P.}~\bibnamefont{Phillips}},
  \bibinfo{journal}{Phys. Rev. B} \textbf{\bibinfo{volume}{79}},
  \bibinfo{pages}{115116} (\bibinfo{year}{2009}).

\bibitem[{\citenamefont{Lee et~al.}(2006{\natexlab{a}})\citenamefont{Lee,
  Nagaosa, and Wen}}]{leenag06}
\bibinfo{author}{\bibfnamefont{P.~A.} \bibnamefont{Lee}},
  \bibinfo{author}{\bibfnamefont{N.}~\bibnamefont{Nagaosa}}, \bibnamefont{and}
  \bibinfo{author}{\bibfnamefont{X.-G.} \bibnamefont{Wen}},
  \bibinfo{journal}{Rev. Mod. Phys.} \textbf{\bibinfo{volume}{78}},
  \bibinfo{pages}{17} (\bibinfo{year}{2006}{\natexlab{a}}).

\bibitem[{\citenamefont{Marzari and Vanderbilt}(1997)}]{mar97}
\bibinfo{author}{\bibfnamefont{N.}~\bibnamefont{Marzari}} \bibnamefont{and}
  \bibinfo{author}{\bibfnamefont{D.}~\bibnamefont{Vanderbilt}},
  \bibinfo{journal}{Phys. Rev. B} \textbf{\bibinfo{volume}{56}},
  \bibinfo{pages}{12847} (\bibinfo{year}{1997}).

\bibitem[{\citenamefont{Souza et~al.}(2001)\citenamefont{Souza, Marzari, and
  Vanderbilt}}]{sou01}
\bibinfo{author}{\bibfnamefont{I.}~\bibnamefont{Souza}},
  \bibinfo{author}{\bibfnamefont{N.}~\bibnamefont{Marzari}}, \bibnamefont{and}
  \bibinfo{author}{\bibfnamefont{D.}~\bibnamefont{Vanderbilt}},
  \bibinfo{journal}{Phys. Rev. B} \textbf{\bibinfo{volume}{65}},
  \bibinfo{pages}{035109} (\bibinfo{year}{2001}).

\bibitem[{\citenamefont{Meyer et~al.}(unpublished)\citenamefont{Meyer,
  Els\"{a}sser, Lechermann, and F\"{a}hnle}}]{mbpp_code}
\bibinfo{author}{\bibfnamefont{B.}~\bibnamefont{Meyer}},
  \bibinfo{author}{\bibfnamefont{C.}~\bibnamefont{Els\"{a}sser}},
  \bibinfo{author}{\bibfnamefont{F.}~\bibnamefont{Lechermann}},
  \bibnamefont{and}
  \bibinfo{author}{\bibfnamefont{M.}~\bibnamefont{F\"{a}hnle}},
  \emph{\bibinfo{title}{FORTRAN 90 Program for Mixed-Basis-Pseudopotential
  Calculations for Crystals}}, \bibinfo{organization}{Max-Planck-Institut
  f\"{u}r Metallforschung, Stuttgart} (\bibinfo{year}{unpublished}).

\bibitem[{\citenamefont{Louie et~al.}(1979)\citenamefont{Louie, Ho, and
  Cohen}}]{lou79}
\bibinfo{author}{\bibfnamefont{S.~G.} \bibnamefont{Louie}},
  \bibinfo{author}{\bibfnamefont{K.~M.} \bibnamefont{Ho}}, \bibnamefont{and}
  \bibinfo{author}{\bibfnamefont{M.~L.} \bibnamefont{Cohen}},
  \bibinfo{journal}{Phys. Rev. B} \textbf{\bibinfo{volume}{19}},
  \bibinfo{pages}{1774} (\bibinfo{year}{1979}).

\bibitem[{\citenamefont{Huang et~al.}(2004)\citenamefont{Huang, Foo, Lynn,
  Zandbergen, Lawes, Wang, Toby, Ramirez, Ong, and Cava}}]{hua04}
\bibinfo{author}{\bibfnamefont{Q.}~\bibnamefont{Huang}},
  \bibinfo{author}{\bibfnamefont{M.~L.} \bibnamefont{Foo}},
  \bibinfo{author}{\bibfnamefont{J.~W.} \bibnamefont{Lynn}},
  \bibinfo{author}{\bibfnamefont{H.~W.} \bibnamefont{Zandbergen}},
  \bibinfo{author}{\bibfnamefont{G.}~\bibnamefont{Lawes}},
  \bibinfo{author}{\bibfnamefont{Y.}~\bibnamefont{Wang}},
  \bibinfo{author}{\bibfnamefont{B.~H.} \bibnamefont{Toby}},
  \bibinfo{author}{\bibfnamefont{A.~P.} \bibnamefont{Ramirez}},
  \bibinfo{author}{\bibfnamefont{N.~P.} \bibnamefont{Ong}}, \bibnamefont{and}
  \bibinfo{author}{\bibfnamefont{R.~J.} \bibnamefont{Cava}},
  \bibinfo{journal}{J. Phys.: Condens. Matter} \textbf{\bibinfo{volume}{16}},
  \bibinfo{pages}{5803} (\bibinfo{year}{2004}).

\bibitem[{\citenamefont{Huang et~al.}(2009)\citenamefont{Huang, Shu, Chu, Kuo,
  Lee, Sheu, and Chou}}]{hua09}
\bibinfo{author}{\bibfnamefont{F.-T.} \bibnamefont{Huang}},
  \bibinfo{author}{\bibfnamefont{G.~J.} \bibnamefont{Shu}},
  \bibinfo{author}{\bibfnamefont{M.-W.} \bibnamefont{Chu}},
  \bibinfo{author}{\bibfnamefont{Y.~K.} \bibnamefont{Kuo}},
  \bibinfo{author}{\bibfnamefont{W.~L.} \bibnamefont{Lee}},
  \bibinfo{author}{\bibfnamefont{H.~S.} \bibnamefont{Sheu}}, \bibnamefont{and}
  \bibinfo{author}{\bibfnamefont{F.~C.} \bibnamefont{Chou}},
  \bibinfo{journal}{Phys. Rev. B} \textbf{\bibinfo{volume}{80}},
  \bibinfo{pages}{144113} (\bibinfo{year}{2009}).

\bibitem[{\citenamefont{Wen et~al.}(2010)\citenamefont{Wen, R\"{u}egg, Wang,
  and Fiete}}]{wen10}
\bibinfo{author}{\bibfnamefont{J.}~\bibnamefont{Wen}},
  \bibinfo{author}{\bibfnamefont{A.}~\bibnamefont{R\"{u}egg}},
  \bibinfo{author}{\bibfnamefont{C.-C.~J.} \bibnamefont{Wang}},
  \bibnamefont{and} \bibinfo{author}{\bibfnamefont{G.~A.} \bibnamefont{Fiete}},
  \bibinfo{journal}{arXiv:1005.4061}  (\bibinfo{year}{2010}).

\bibitem[{\citenamefont{Lichtenstein et~al.}(2003)\citenamefont{Lichtenstein,
  Katsnelson, and Kotliar}}]{lic03}
\bibinfo{author}{\bibfnamefont{A.}~\bibnamefont{Lichtenstein}},
  \bibinfo{author}{\bibfnamefont{M.}~\bibnamefont{Katsnelson}},
  \bibnamefont{and} \bibinfo{author}{\bibfnamefont{G.}~\bibnamefont{Kotliar}},
  \emph{\bibinfo{title}{in Electron Correlations and Materials Properties 2}}
  (\bibinfo{publisher}{Kluwer Academic}, \bibinfo{year}{2003}).

\bibitem[{\citenamefont{Biroli et~al.}(2004)\citenamefont{Biroli, Parcollet,
  and Kotliar}}]{bir04}
\bibinfo{author}{\bibfnamefont{G.}~\bibnamefont{Biroli}},
  \bibinfo{author}{\bibfnamefont{O.}~\bibnamefont{Parcollet}},
  \bibnamefont{and} \bibinfo{author}{\bibfnamefont{G.}~\bibnamefont{Kotliar}},
  \bibinfo{journal}{Phys. Rev. B} \textbf{\bibinfo{volume}{69}},
  \bibinfo{pages}{205108} (\bibinfo{year}{2004}).

\bibitem[{\citenamefont{Maier et~al.}(2005)\citenamefont{Maier, Jarrell,
  Pruschke, and Hettler}}]{mai05}
\bibinfo{author}{\bibfnamefont{T.}~\bibnamefont{Maier}},
  \bibinfo{author}{\bibfnamefont{M.}~\bibnamefont{Jarrell}},
  \bibinfo{author}{\bibfnamefont{T.}~\bibnamefont{Pruschke}}, \bibnamefont{and}
  \bibinfo{author}{\bibfnamefont{M.~H.} \bibnamefont{Hettler}},
  \bibinfo{journal}{Rev. Mod. Phys.} \textbf{\bibinfo{volume}{77}},
  \bibinfo{pages}{1027} (\bibinfo{year}{2005}).

\bibitem[{\citenamefont{Ferrero et~al.}(2009)\citenamefont{Ferrero, Cornaglia,
  Leo, Parcollet, Kotliar, and Georges}}]{fer09}
\bibinfo{author}{\bibfnamefont{M.}~\bibnamefont{Ferrero}},
  \bibinfo{author}{\bibfnamefont{P.~S.} \bibnamefont{Cornaglia}},
  \bibinfo{author}{\bibfnamefont{L.~D.} \bibnamefont{Leo}},
  \bibinfo{author}{\bibfnamefont{O.}~\bibnamefont{Parcollet}},
  \bibinfo{author}{\bibfnamefont{G.}~\bibnamefont{Kotliar}}, \bibnamefont{and}
  \bibinfo{author}{\bibfnamefont{A.}~\bibnamefont{Georges}},
  \bibinfo{journal}{Europhys. Lett.} \textbf{\bibinfo{volume}{85}},
  \bibinfo{pages}{57009} (\bibinfo{year}{2009}).

\bibitem[{\citenamefont{Capone et~al.}(2001)\citenamefont{Capone, Capriotti,
  Becca, and Caprara}}]{cap01}
\bibinfo{author}{\bibfnamefont{M.}~\bibnamefont{Capone}},
  \bibinfo{author}{\bibfnamefont{L.}~\bibnamefont{Capriotti}},
  \bibinfo{author}{\bibfnamefont{F.}~\bibnamefont{Becca}}, \bibnamefont{and}
  \bibinfo{author}{\bibfnamefont{S.}~\bibnamefont{Caprara}},
  \bibinfo{journal}{Phys. Rev. B} \textbf{\bibinfo{volume}{63}},
  \bibinfo{pages}{085104} (\bibinfo{year}{2001}).

\bibitem[{\citenamefont{Kyung}(2007)}]{kyu07}
\bibinfo{author}{\bibfnamefont{B.}~\bibnamefont{Kyung}},
  \bibinfo{journal}{Phys. Rev. B} \textbf{\bibinfo{volume}{75}},
  \bibinfo{pages}{033102} (\bibinfo{year}{2007}).

\bibitem[{\citenamefont{Mukhamedshin et~al.}(2004)\citenamefont{Mukhamedshin,
  Alloul, Collin, and Blanchard}}]{muk04}
\bibinfo{author}{\bibfnamefont{I.~R.} \bibnamefont{Mukhamedshin}},
  \bibinfo{author}{\bibfnamefont{H.}~\bibnamefont{Alloul}},
  \bibinfo{author}{\bibfnamefont{G.}~\bibnamefont{Collin}}, \bibnamefont{and}
  \bibinfo{author}{\bibfnamefont{N.}~\bibnamefont{Blanchard}},
  \bibinfo{journal}{Phys. Rev. Lett.} \textbf{\bibinfo{volume}{93}},
  \bibinfo{pages}{167601} (\bibinfo{year}{2004}).

\bibitem[{\citenamefont{Sakurai et~al.}(2006)\citenamefont{Sakurai, Tsujii, and
  Takayama-Muromachi}}]{sak06}
\bibinfo{author}{\bibfnamefont{H.}~\bibnamefont{Sakurai}},
  \bibinfo{author}{\bibfnamefont{N.}~\bibnamefont{Tsujii}}, \bibnamefont{and}
  \bibinfo{author}{\bibfnamefont{E.}~\bibnamefont{Takayama-Muromachi}},
  \bibinfo{journal}{Physica B} \textbf{\bibinfo{volume}{378}}
  (\bibinfo{year}{2006}).

\bibitem[{\citenamefont{Balicas et~al.}(2008)\citenamefont{Balicas, Jo, Shu,
  Chou, and Lee}}]{bal08}
\bibinfo{author}{\bibfnamefont{L.}~\bibnamefont{Balicas}},
  \bibinfo{author}{\bibfnamefont{Y.~J.} \bibnamefont{Jo}},
  \bibinfo{author}{\bibfnamefont{G.~J.} \bibnamefont{Shu}},
  \bibinfo{author}{\bibfnamefont{F.~C.} \bibnamefont{Chou}}, \bibnamefont{and}
  \bibinfo{author}{\bibfnamefont{P.~A.} \bibnamefont{Lee}},
  \bibinfo{journal}{Phys. Rev. Lett.} \textbf{\bibinfo{volume}{100}},
  \bibinfo{pages}{126405} (\bibinfo{year}{2008}).

\bibitem[{\citenamefont{Zorkovsk{\'a} et~al.}(2010)\citenamefont{Zorkovsk{\'a},
  Baran, Kajakov{\'a}, Feher, Sebek, Santav{\'a}, Lin, and Peng}}]{zor10}
\bibinfo{author}{\bibfnamefont{A.}~\bibnamefont{Zorkovsk{\'a}}},
  \bibinfo{author}{\bibfnamefont{A.}~\bibnamefont{Baran}},
  \bibinfo{author}{\bibfnamefont{M.}~\bibnamefont{Kajakov{\'a}}},
  \bibinfo{author}{\bibfnamefont{A.}~\bibnamefont{Feher}},
  \bibinfo{author}{\bibfnamefont{J.}~\bibnamefont{Sebek}},
  \bibinfo{author}{\bibfnamefont{E.}~\bibnamefont{Santav{\'a}}},
  \bibinfo{author}{\bibfnamefont{C.~T.} \bibnamefont{Lin}}, \bibnamefont{and}
  \bibinfo{author}{\bibfnamefont{J.~B.} \bibnamefont{Peng}},
  \bibinfo{journal}{Phys. Stat. Sol. (b)} \textbf{\bibinfo{volume}{247}},
  \bibinfo{pages}{665} (\bibinfo{year}{2010}).

\bibitem[{\citenamefont{Lee et~al.}(2006{\natexlab{b}})\citenamefont{Lee,
  Viciu, Li, Wang, Foo, Watauchi, Pascal, Cava, and Ong}}]{lee06}
\bibinfo{author}{\bibfnamefont{M.}~\bibnamefont{Lee}},
  \bibinfo{author}{\bibfnamefont{L.}~\bibnamefont{Viciu}},
  \bibinfo{author}{\bibfnamefont{L.}~\bibnamefont{Li}},
  \bibinfo{author}{\bibfnamefont{Y.}~\bibnamefont{Wang}},
  \bibinfo{author}{\bibfnamefont{M.~L.} \bibnamefont{Foo}},
  \bibinfo{author}{\bibfnamefont{S.}~\bibnamefont{Watauchi}},
  \bibinfo{author}{\bibfnamefont{R.~A.} \bibnamefont{Pascal}},
  \bibinfo{author}{\bibfnamefont{R.~J.} \bibnamefont{Cava}}, \bibnamefont{and}
  \bibinfo{author}{\bibfnamefont{N.~P.} \bibnamefont{Ong}},
  \bibinfo{journal}{Nat. Mat.} \textbf{\bibinfo{volume}{5}},
  \bibinfo{pages}{537} (\bibinfo{year}{2006}{\natexlab{b}}).

\bibitem[{\citenamefont{Motrunich and Lee}(2004)}]{mot04}
\bibinfo{author}{\bibfnamefont{O.~I.} \bibnamefont{Motrunich}}
  \bibnamefont{and} \bibinfo{author}{\bibfnamefont{P.~A.} \bibnamefont{Lee}},
  \bibinfo{journal}{Phys. Rev. B} \textbf{\bibinfo{volume}{69}},
  \bibinfo{pages}{214516} (\bibinfo{year}{2004}).

\bibitem[{\citenamefont{Hassan et~al.}(2007)\citenamefont{Hassan, de~Medici,
  and Tremblay}}]{has07}
\bibinfo{author}{\bibfnamefont{S.~R.} \bibnamefont{Hassan}},
  \bibinfo{author}{\bibfnamefont{L.}~\bibnamefont{de~Medici}},
  \bibnamefont{and} \bibinfo{author}{\bibfnamefont{A.-M.~S.}
  \bibnamefont{Tremblay}}, \bibinfo{journal}{Phys. Rev. B}
  \textbf{\bibinfo{volume}{76}}, \bibinfo{pages}{144420}
  (\bibinfo{year}{2007}).

\bibitem[{\citenamefont{Bejas et~al.}(2008)\citenamefont{Bejas, Greco,
  Muramatsu, and Foussats}}]{bej08}
\bibinfo{author}{\bibfnamefont{M.}~\bibnamefont{Bejas}},
  \bibinfo{author}{\bibfnamefont{A.}~\bibnamefont{Greco}},
  \bibinfo{author}{\bibfnamefont{A.}~\bibnamefont{Muramatsu}},
  \bibnamefont{and} \bibinfo{author}{\bibfnamefont{A.}~\bibnamefont{Foussats}},
  \bibinfo{journal}{Phys. Rev. B} \textbf{\bibinfo{volume}{77}},
  \bibinfo{pages}{075131} (\bibinfo{year}{2008}).

\bibitem[{\citenamefont{Qian et~al.}(2006)\citenamefont{Qian, Hsieh, Wray,
  Chuang, Fedorov, Wu, Lue, Wang, Viciu, Cava et~al.}}]{qia06}
\bibinfo{author}{\bibfnamefont{D.}~\bibnamefont{Qian}},
  \bibinfo{author}{\bibfnamefont{D.}~\bibnamefont{Hsieh}},
  \bibinfo{author}{\bibfnamefont{L.}~\bibnamefont{Wray}},
  \bibinfo{author}{\bibfnamefont{Y.-D.} \bibnamefont{Chuang}},
  \bibinfo{author}{\bibfnamefont{A.}~\bibnamefont{Fedorov}},
  \bibinfo{author}{\bibfnamefont{D.}~\bibnamefont{Wu}},
  \bibinfo{author}{\bibfnamefont{J.~L.} \bibnamefont{Lue}},
  \bibinfo{author}{\bibfnamefont{N.~L.} \bibnamefont{Wang}},
  \bibinfo{author}{\bibfnamefont{L.}~\bibnamefont{Viciu}},
  \bibinfo{author}{\bibfnamefont{R.~J.} \bibnamefont{Cava}},
  \bibnamefont{et~al.}, \bibinfo{journal}{Phys. Rev. Lett.}
  \textbf{\bibinfo{volume}{96}}, \bibinfo{pages}{216405}
  (\bibinfo{year}{2006}).

\bibitem[{\citenamefont{Brouet et~al.}(2007)\citenamefont{Brouet, Nicolaou,
  Zacchigna, Tejeda, Patthey, H{\'e}bert, Kobayashi, Muguerra, and
  Grebille}}]{bro07}
\bibinfo{author}{\bibfnamefont{V.}~\bibnamefont{Brouet}},
  \bibinfo{author}{\bibfnamefont{A.}~\bibnamefont{Nicolaou}},
  \bibinfo{author}{\bibfnamefont{M.}~\bibnamefont{Zacchigna}},
  \bibinfo{author}{\bibfnamefont{A.}~\bibnamefont{Tejeda}},
  \bibinfo{author}{\bibfnamefont{L.}~\bibnamefont{Patthey}},
  \bibinfo{author}{\bibfnamefont{S.}~\bibnamefont{H{\'e}bert}},
  \bibinfo{author}{\bibfnamefont{W.}~\bibnamefont{Kobayashi}},
  \bibinfo{author}{\bibfnamefont{H.}~\bibnamefont{Muguerra}}, \bibnamefont{and}
  \bibinfo{author}{\bibfnamefont{D.}~\bibnamefont{Grebille}},
  \bibinfo{journal}{Phys. Rev. B} \textbf{\bibinfo{volume}{76}},
  \bibinfo{pages}{100403} (\bibinfo{year}{2007}).

\bibitem[{\citenamefont{Nicolaou et~al.}(2010)\citenamefont{Nicolaou, Brouet,
  Zacchigna, Vobornik, Tejeda, Taleb-Ibrahimi, F{\`e}vre, Bertran, H{\'e}bert,
  Muguerra et~al.}}]{nic10}
\bibinfo{author}{\bibfnamefont{A.}~\bibnamefont{Nicolaou}},
  \bibinfo{author}{\bibfnamefont{V.}~\bibnamefont{Brouet}},
  \bibinfo{author}{\bibfnamefont{M.}~\bibnamefont{Zacchigna}},
  \bibinfo{author}{\bibfnamefont{I.}~\bibnamefont{Vobornik}},
  \bibinfo{author}{\bibfnamefont{A.}~\bibnamefont{Tejeda}},
  \bibinfo{author}{\bibfnamefont{A.}~\bibnamefont{Taleb-Ibrahimi}},
  \bibinfo{author}{\bibfnamefont{P.~L.} \bibnamefont{F{\`e}vre}},
  \bibinfo{author}{\bibfnamefont{F.}~\bibnamefont{Bertran}},
  \bibinfo{author}{\bibfnamefont{S.}~\bibnamefont{H{\'e}bert}},
  \bibinfo{author}{\bibfnamefont{H.}~\bibnamefont{Muguerra}},
  \bibnamefont{et~al.}, \bibinfo{journal}{Phys. Rev. Lett.}
  \textbf{\bibinfo{volume}{104}}, \bibinfo{pages}{056403}
  (\bibinfo{year}{2010}).

\bibitem[{\citenamefont{Bobroff et~al.}(2007)\citenamefont{Bobroff, H\'{e}bert,
  Lang, Mendels, Pelloquin, and Maignan}}]{bob07}
\bibinfo{author}{\bibfnamefont{J.}~\bibnamefont{Bobroff}},
  \bibinfo{author}{\bibfnamefont{S.}~\bibnamefont{H\'{e}bert}},
  \bibinfo{author}{\bibfnamefont{G.}~\bibnamefont{Lang}},
  \bibinfo{author}{\bibfnamefont{P.}~\bibnamefont{Mendels}},
  \bibinfo{author}{\bibfnamefont{D.}~\bibnamefont{Pelloquin}},
  \bibnamefont{and} \bibinfo{author}{\bibfnamefont{A.}~\bibnamefont{Maignan}},
  \bibinfo{journal}{Phys. Rev. B} \textbf{\bibinfo{volume}{76}},
  \bibinfo{pages}{100407} (\bibinfo{year}{2007}).

\bibitem[{\citenamefont{Nagaoka}(1966)}]{nag66}
\bibinfo{author}{\bibfnamefont{Y.}~\bibnamefont{Nagaoka}},
  \bibinfo{journal}{Phys. Rev.} \textbf{\bibinfo{volume}{147}},
  \bibinfo{pages}{392} (\bibinfo{year}{1966}).

\bibitem[{\citenamefont{Kanamori}(1963)}]{kan63}
\bibinfo{author}{\bibfnamefont{J.}~\bibnamefont{Kanamori}},
  \bibinfo{journal}{Progr. Theor. Phys.} \textbf{\bibinfo{volume}{30}},
  \bibinfo{pages}{275} (\bibinfo{year}{1963}).

\bibitem[{\citenamefont{Tasaki}(1998)}]{tas98}
\bibinfo{author}{\bibfnamefont{H.}~\bibnamefont{Tasaki}}, \bibinfo{journal}{J.
  Phys.: Condens. Matter} \textbf{\bibinfo{volume}{10}}, \bibinfo{pages}{4353}
  (\bibinfo{year}{1998}).

\bibitem[{\citenamefont{Penc et~al.}(1996)\citenamefont{Penc, Shiba, Mila, and
  Tsukagoshi}}]{pen96}
\bibinfo{author}{\bibfnamefont{K.}~\bibnamefont{Penc}},
  \bibinfo{author}{\bibfnamefont{H.}~\bibnamefont{Shiba}},
  \bibinfo{author}{\bibfnamefont{F.}~\bibnamefont{Mila}}, \bibnamefont{and}
  \bibinfo{author}{\bibfnamefont{T.}~\bibnamefont{Tsukagoshi}},
  \bibinfo{journal}{Phys. Rev. B} \textbf{\bibinfo{volume}{54}},
  \bibinfo{pages}{4056} (\bibinfo{year}{1996}).

\bibitem[{\citenamefont{Merino et~al.}(2009)\citenamefont{Merino, McKenzie, and
  Powell}}]{mer09}
\bibinfo{author}{\bibfnamefont{J.}~\bibnamefont{Merino}},
  \bibinfo{author}{\bibfnamefont{R.~H.} \bibnamefont{McKenzie}},
  \bibnamefont{and} \bibinfo{author}{\bibfnamefont{B.~J.}
  \bibnamefont{Powell}}, \bibinfo{journal}{Phys. Rev. B}
  \textbf{\bibinfo{volume}{80}}, \bibinfo{pages}{045116}
  (\bibinfo{year}{2009}).

\end{thebibliography}

\end{document}